\documentclass[12pt]{article}
\pdfoutput=1
\usepackage{amssymb}
\usepackage{graphicx}
\usepackage{hyperref}
\usepackage{bm}

\def\v{{\bm v}}
\def\x{{\bm x}}
\def\E{{\bm E}}
\def\B{{\bm B}}
\def\me{m_e}

\def\hp{\hat\phi}
\def\tf{\tilde f}
\def\tF{\tilde F}
\def\tn{\tilde n}

\def\tL{\tilde L}

\ifx\affiliation\undefined 
\def\affiliation#1{\date{\normalsize #1\\ \today}}
\def\Real{{\rm Re}}\def\Imag{{\rm Im}}
\usepackage[margin=1in]{geometry}
\def\citep{\cite}                   
\else 
\fi

\title{Transverse instability of electron phase-space holes in
  multi-dimensional Maxwellian plasmas}

\author{I H Hutchinson}

\affiliation{Plasma Science and Fusion Center, MIT, Cambridge, MA
  02139, USA}

\begin{document}
\maketitle

\begin{abstract}
  The stability of an initially one-dimensional electron hole to
  perturbations varying sinusoidally transverse to its trapping
  direction is analysed in detail. It is shown that the expected
  low-frequency eigenmode of the linearized Vlasov-Poisson system
  consists of a shift-mode, proportional to the gradient of the
  equilibrium potential. The resulting dispersion relation is that the
  total jetting force exerted by a perturbed hole on the particles
  balances the electric restoring tension of the hole. The tension is
  quantitatively small and can often be ignored. The particle force is
  expressed as integrals of equilibrium parameters over the hole and
  is shown at low frequency to be exactly equal to what has recently
  been found (by different analysis) to express `kinematic' hole
  momentum conservation. The mechanism of instability has nothing to
  do with the previously hypothesized transverse electron
  focusing. The unmagnetized growth rate $\gamma(k)$ is found
  numerically and is in excellent agreement with recent kinematic
  estimates. Magnetic field stabilization of the transverse mode is
  also evaluated. The resulting stability boundary for Maxwellian
  holes is in reasonable agreement with previously published criteria
  based on particle simulation. It arises from a change of trapped
  force sign across the resonance between bounce and cyclotron
  frequencies.
\end{abstract}

\section{Introduction}

An electron hole in a plasma is a solitary BGK
mode \citep{Bernstein1957} consisting of a positive potential peak
self-consistently maintained by phase-space density deficit of trapped
electrons \citep{Turikov1984,Schamel1986a,Eliasson2006,Hutchinson2017}. In
one dimension, electron holes are the normal nonlinear consequence of
growth of an electrostatic instability driven by an unstable electron
velocity distribution, for example a two-stream instability. However,
it has been known since the earliest simulations \citep{Morse1969} that
in multiple dimensions electron holes either do not form or quickly
break up by instability permitted by the additional dimensions; this
is referred to as the transverse instability, and observed in many
simulations
since \citep{Mottez1997,Miyake1998a,Goldman1999,Oppenheim1999,Muschietti2000,Oppenheim2001b,Singh2001,Lu2008}. Despite
the obvious significance of this instability, since it determines the
prevalence and fate of electron holes, satisfactory identification
of its mechanism and rigorous analytical determination of its growth
rate and threshold have until now been lacking. The current paper now
presents this analysis.

Satellite measurements with high time-resolution during the past 20
years have established
\citep{Matsumoto1994,Ergun1998,Bale1998,Mangeney1999,Pickett2008,Andersson2009,Wilson2010,Malaspina2013,Malaspina2014,Vasko2015,Mozer2016}
that electron holes are widely present in a variety of space plasma
regions. They are likely present also, though harder to resolve and
observe, in many laboratory plasmas. Therefore it is increasingly
pressing to develop a fundamental understanding of their persistence
and stability. 

Particle in cell (and some Vlasov continuum)
simulations have been important in mapping out the phenomena.  However,
such simulations have not correctly identified the underlying
instability mechanism.
\begin{figure}
  \centering
  \includegraphics[width=.5\hsize]{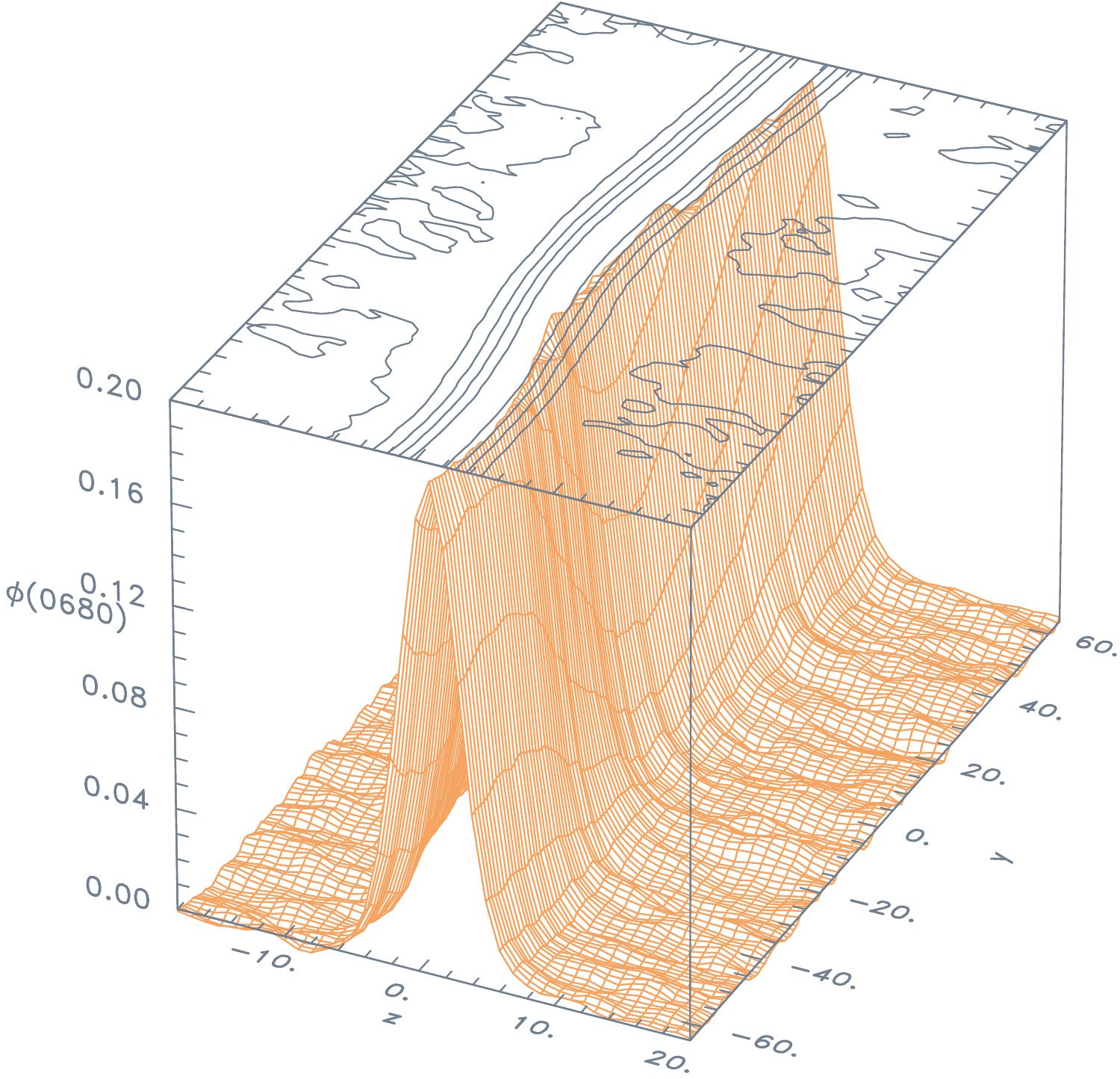}
  \caption{Two-dimensional $(z,y)$ rendering and contours of potential $\phi$
    of an initially one-dimensional hole in the early stages of a
    transverse instability in a PIC simulation.}
  \label{fig:holekink}
\end{figure}
Fig.\ \ref{fig:holekink} illustrates an example of a simulation,
illustrating a hole and a growing kink in it.
The natural idealization of the problem, concentrating on the
stability properties rather than on hole formation, is to begin with a
pre-formed hole (slab) equilibrium that varies in only one
dimension. Simulations of this type have provided valuable insight
\citep{Muschietti1999,Muschietti2000,Wu2010}, and identified the
stabilizing effect of magnetic field. However the transverse
`focusing' mechanism, which they hypothesized causes the instability, is not
confirmed either by subsequent simulations \citep{hutchinson18} or by the
present analysis. 

Analytic approaches to solving this linear stability problem have
previously been at best only partly successful
\citep{Schamel1982,Schamel1987,Jovanovic2002}. A key difficulty lies
in identifying the appropriate eigenmode structure in the direction of
equilibrium non-uniformity \citep{Lewis1979}. Previous unmagnetized
analysis \citep{Schamel1982} made the inappropriate choice to address
symmetric eigenmodes. It is shown here that, in agreement with
simulations, the relevant eigenmode is predominantly
\emph{antisymmetric}. Another difficulty is to solve the linearized
Vlasov equation by integration along characteristics in regimes that
do not yield to simple expansions, in part because of
non-uniformity. Previous analyses \citep{Schamel1982,Jovanovic2002}
have used expansion in inverse powers of frequency, which is
inappropriate for these slowly (and purely) growing modes. The present
study instead solves the Vlasov equation by numerical integration at
finite complex frequency, showing excellent agreement at small
$\omega$ with a better motivated analytic approximation. A previous
study of magnetized holes \citep{Jovanovic2002} was, because of its
expansions, limited to cases where the wave or cyclotron frequencies
are bigger than the trapped-particles' highest bounce frequency, yet
concluded there was instability on the basis of a wave resonance
$\omega+m\Omega\to 0$.  The present study shows that the low-$k$
transverse instability is \emph{stabilized} when the cyclotron
frequency exceeds the bounce frequency and that the wave frequency is
always small, so there is no such resonance except $m=0$ (which is
effectively the one-dimensional case).  Prior analytic studies
provided no quantitative comparison with simulations. The present
study gives quantitative predictions of the fastest growing mode and
the previously proposed heuristic criterion \citep{Miyake1998a,Muschietti2000}
for stabilization by magnetic field that agree well with simulations.

In view of these contrasts, a rather careful development of the
mathematics is given here to provide rigor in the derivation, even
though some of it has close parallels in the standard analysis of
uniform equilibria. The results we find are in full agreement with a
more descriptive letter recently published by the author
\citep{hutchinson18}. The analysis provides the mathematical
justification for the identification of the instability mechanism as
being a kinematic one, arising from the transfer of momentum from hole
potential to particles \citep{Hutchinson2016,Zhou2016,Zhou2017},
called jetting.

\section{Linearized Vlasov Equation}

Consider a localized positive potential peak equilibrium $\phi_0(z)$,
the electron hole, to be analysed for linearized electrostatic
perturbations. We address only the electron dynamics, taking the ions
to be a uniform immobile background. We must retain full-wave
treatment in the $z$-direction, but can Fourier analyze the transverse
variation taking a transverse wave-number $k$, without loss of
generality in the $y$-direction. So the first-order potential
perturbation is harmonic
\begin{equation}
  \label{eq:phiexp}
\phi_1(\x,t)=\hp(z)\exp i(ky-\omega t).  
\end{equation}
Vlasov's equation is 
\begin{equation}
  \label{eq:Vlasov}
  \left({\partial \over \partial t} +\v\cdot\nabla + {q_e\over \me}
      (\E+\v\wedge\B)\cdot\nabla_v\right) f =0.
\end{equation}
When linearized in an
electrostatic approximation it becomes
\begin{equation}
  \label{eq:linVlas} {df_1\over dt}\equiv
  \left({\partial \over \partial t} +\v\cdot\nabla + {q_e\over m_e}
    (\E_0+\v\wedge\B_0)\cdot\nabla_v\right) f_1=
  - {q_e\over m_e}\E_1\cdot\nabla_v f_0,
\end{equation}
which is the rate of change of $f_1$ along the unperturbed orbit in
phase space. And
\begin{equation}
  \label{eq:E1}
  {\E}_1=-\nabla\phi_1  =- \left({\partial\hp\over\partial z} 
    \hat{\bm z} +ik\hp
    \hat{\bm y}\right) \exp i(ky-\omega t).
\end{equation}
Integrating eq.\ \ref{eq:linVlas}, we get the perturbed distribution:
\begin{equation}
  \label{eq:f1init}
  f_1(\v,\x,t) =
  {q_e\over\me}\int_{-\infty}^t\left({d\hp\over dz}
    {\partial f_0\over\partial v_z} +ik\hp{\partial f_0\over \partial
      v_y}\right) \exp i(ky-\omega\tau) d\tau,
\end{equation}
where the integrand is to be evaluated on the unperturbed orbit, i.e.\
at $\x(\tau)$, $\v(\tau)$, which is the characteristic of the
linearized equation.

The equilibrium distribution is a function of the constants of the
unperturbed motion, which are the total parallel energy
$W_\parallel=q_e\phi+v_\parallel^2/2$ (with $z$ the `parallel'
direction) and the perpendicular kinetic
energy $W_\perp\equiv \me v_\perp^2/2$. If a uniform background
magnetic field, $B_0\hat{\bf z}$ is present, then
$v_\perp^2=v_x^2+v_y^2$, but if $B=0$ we can interpret $v_\perp$ as
$v_y$, and ignore $v_x$. So write
\begin{equation}
  \label{eq:energyforms}
  {\partial f_0\over \partial v_z}=  
  \me v_z{\partial f_0\over \partial W_\parallel}\quad,\quad
  {\partial f_0\over \partial v_y}=mv_y{\partial f_0\over \partial W_\perp}.
\end{equation}
And 
since
$v_z{d\hp\over dz}
={d\hp\over d\tau}$,
\begin{equation}
  \label{eq:f1energy}
  f_1(\v,\x,t) =
  q_e\int_{-\infty}^t\left(
    {d\hp\over d\tau}
    {\partial f_0\over\partial W_\parallel} +ikv_y\hp{\partial f_0\over \partial
      W_\perp}\right) \exp i(ky-\omega\tau) d\tau.
\end{equation}
The first term can be integrated by parts to give
\begin{equation}
  \label{eq:f1parts}
  f_1(t) = q_e\phi_1(t){\partial f_0\over\partial W_\parallel}+
  q_e\int_{-\infty}^t\left(i(\omega-kv_y)
    {\partial f_0\over\partial W_\parallel} +ikv_y{\partial f_0\over \partial
      W_\perp}\right)\hp {\rm e}^{i(ky-\omega\tau)} d\tau,    
\end{equation}
assuming that $\phi_1(t=-\infty)=0$.
The leading term is then the `adiabatic' response arising from the
variation of a distribution that remains a constant function of total parallel
energy $W_\parallel$ in response to a potential
perturbation giving $dW_\parallel=q_e\phi_1$. The integral term is the
non-adiabatic response, which we shall denote $\tf$. 

It is worth noticing that when the unperturbed distribution has a
separable isotropic form in energy, for example on all passing orbits
when the background is (isotropic) Maxwellian, the $kv_y$ terms
disappear because
\begin{equation}
  \label{eq:isomax}
  -{\partial f_0\over\partial W_\parallel} +{\partial f_0\over \partial
      W_\perp}=0.
\end{equation}
But this happens only if the isotropy is in the hole's rest frame.

\section{The unstable shift eigenmode}

\subsection{Eigen-analysis of the Poisson-Vlasov system}

The perturbed Poisson equation (leaving the factor
${\rm e}^{i(ky-\omega t)}$ implicit) is
\begin{equation}
  \label{eq:Poisson}
  \nabla^2\phi_1 = {d^2 \hp\over d z^2} -k^2\hp=
  -{q_e\over \epsilon_0} \int f_{\parallel1} dv_z=
  -{q_e\over \epsilon_0}\left( \hp {d n_0\over d\phi_0} 
  +\int \tf_{\parallel}dv_z\right).
\end{equation}
The structure of the linear stability problem is then that we can
regard the quantity $\tn =\int \tf_\parallel dv_z$ as a linear
functional of $\hp$ determined by the solution of the linearized
Vlasov equation. The Poisson equation for given $k$ is then a
generalized eigen-problem \citep{Lewis1979}, of which the complex $\omega$ is
effectively the eigenvalue and the spatial shape of $\hp$ is the
eigenmode. The non-uniformity of the equilibrium means that the
eigenmode structure is of course \emph{not} a Fourier
mode. Determining all the eigenmodes and eigenvalues exactly is
difficult, in part because $\tn$ is an integral, rather than
differential, functional.

However, we are not attempting here to identify every mode; only the
mechanism and growth rate of a specific type of perturbation observed
to occur in simulations, namely the transverse instability. Therefore
we concentrate on obtaining a good estimate of the eigenvalue,
assuming we have a fair approximation to what the eigenmode shape
is. For this purpose, we presuppose what characteristics of the
eigenmode the simulations indicate, namely that it is present for
small finite values of $k$ (long transverse wavelength) and the
magnitude of its complex frequency is small. To lowest order, then,
the eigenmode equation can omit the terms $k^2\phi$ and $\tn$, in
which case it must approximately satisfy
\begin{equation}
  \label{eq:eigen0}
  {d^2 \hp\over d z^2} =
  -{q_e\over \epsilon_0}{d n_0\over d\phi_0}\hp = \hp {d\over d\phi_0}\left(d^2\phi_0\over dz^2\right) .
\end{equation}
It is easy to verify that this linear equation is satisfied by
$\hp = -\Delta{d \phi_0\over d z}$, a shift-mode, which is sometimes
in soliton and other literature referred to as the Goldstone mode (see
e.g.\  \citep{Skryabin2002}). It consists of simply a displacement of
the entire hole structure in the $z$-direction, which is obviously a
neighboring equilibrium, requiring no time derivative terms. In so far
as there is a unique solution (modulo a position shift) for a BGK mode
(electron hole) with specified distribution function, the shift
appears to be the unique solution of this adiabatic equation. In any
event, the simulations show that the mode does in fact consist
predominantly of a shift, with some other minor distortions.

We then take advantage of the fact that an eigen-problem
${\cal L}\hp =\lambda {\cal M}\hp$ where ${\cal L}$ and ${\cal M}$ are
self-adjoint operators\footnote{Reference \citep{Schamel1987} shows
  the present problem's operators are self-adjoint under some
  restrictions.}, is equivalent to a variational problem that finds
the extremum of the quotient
$Q=\langle \hp {\cal L}\hp\rangle/\langle \hp {\cal M}\hp\rangle$, and
that the eigenvalue therefore deviates from the quotient only by terms
second order in any deviation of the $\hp$ used in evaluating it from
the exact eigenmode. Applying this approach we substitute the
shift-mode into the full Poisson equation, multiply by $\hp$, and
integrate over space. This process annihilates the adiabatic terms
leaving
\begin{equation}
  \label{eq:quotient}
  \Delta k^2\int \epsilon_0\left(d \phi_0\over d z\right)^2dz =-\int{d \phi_0\over d z} q_e \int
  \tf_{\parallel}dv_z dz.
\end{equation}
The value of $\omega$ that satisfies this equation is
then anticipated to be a good approximation to the eigen-frequency of
the mode. 

\subsection{Force Balance}
A simple physical interpretation of eq.\ (\ref{eq:quotient}) is that
it is the conservation of momentum: balancing the electric field
tension (LHS) with the electric force on the non-adiabatic electron
perturbation (RHS). That makes it an overall approximate `kinematic'
constraint.  Indeed, this interpretation is one heuristic explanation
of the transverse instability mechanism. It is a balance between the
effects of `jetting' --- the momentum transfer to particles because of the
acceleration of the hole \citep{Hutchinson2016} --- (RHS), and the
stabilizing effects of electric field tension (LHS). We shall see
shortly that in most circumstances the jetting is the predominant term
and the stability boundary is approximately where the jetting is
reduced to zero.

We can obtain this equation directly through force considerations
without any immediate assumptions about the perturbation eigenmode
shape, as follows.  Force balance requires the total first-order force
exerted by the electric field to be zero, because the electric field
of a solitary potential structure is incapable of sustaining any net
momentum transfer. Denoting the charge density by $\rho$, this is
$E_0\rho_1$ plus $E_1\rho_0$ spatially integrated over the hole.
These two forces (summed over species) are equal and opposite for a
straight ($k=0$) isolated hole because,
\begin{equation}
  \label{eq:totalforce}
  \int{d\phi_0\over dz}{\rho_1\over \epsilon_0}dz=  
  -\int{d\phi_0\over dz}{d^2\phi_1\over  dz^2}dz=
  -\left[{d\phi_0\over dz}{d\phi_1\over dz}\right]
  +\int{d^2\phi_0\over dz^2}{d\phi_1\over  dz}dz=
  -\int{\rho_0\over\epsilon_0}{d\phi_1\over  dz}dz.
\end{equation}
Integrating by parts the final form of eq.\ (\ref{eq:totalforce})
noting that $\rho_0$ is a function of $\phi_0$, we
find
\begin{equation}
  \label{eq:kinkforcegen}
\int{\rho_0}{d\phi_1\over  dz}dz
  =
  - \int {d\rho_0\over d\phi_0} {d\phi_0\over dz}\phi_1 dz. 
\end{equation}
A hole with $y$-variation ($k\not=0$) has an additional term in
$\rho_1$, arising
from ${\partial^2 \phi_1\over \partial y^2} =-k^2\phi_1$. Including it
we get
\begin{equation}
  \label{eq:forcebal}
   \tF \equiv -\int{d\phi_0\over dz}\left[\rho_1- {d\rho_0\over
     d\phi_0}\phi_1\right]  dz= -\epsilon_0 k^2 \int {d\phi_0\over dz}
 \phi_1 dz
 \equiv F_E.
\end{equation}
The term in the square brackets is manifestly the non-adiabatic charge
density $q_e\tn=q_e\int \tf dv$. Notice that because $\phi_0$ is
symmetric, this equation selects the \emph{antisymmetric} part
of $\phi_1$.  For a shift mode it is equivalent to the form
$\int\phi_1^* \tf d^3vd^3x$ (whose variational Euler equation is the
eigen-problem). The right hand side becomes
$F_E=\Delta k^2 \int\epsilon_0 \left(d\phi_0\over dz\right)^2 dz$,
which can be thought of as arising from the transfer of $z$-momentum in the
$y$-direction via the electric field off-diagonal components
of the Maxwell stress tensor (${\partial\over \partial y} E_yE_z$). It causes the hole to resist
kinking as if it had a tension in the $y$-direction.

\section{Unmagnetized Plasmas}

When there is no magnetic field, the transverse velocity $v_y$ of the
equilibrium (unperturbed) orbit is independent of time, and the $\tau$
dependence in the $\tf$ expression (\ref{eq:f1parts}) resides only in
$\hp(z(\tau))$ and the exponent.  In the $\tau$-integral
$y=y(\tau)= y(t)+v_y(\tau-t)$.  We use time subscripts as shorthand
for arguments ($y(t)\equiv y_t$ etc.) and $\omega'=\omega-kv_y$ to
denote the frequency felt by the $y$-moving particle. Then we can
write
\begin{equation}
  \label{eq:tf0}
  \tf = iq_e\left((\omega-kv_y)
    {\partial f_0\over\partial W_\parallel} +kv_y{\partial f_0\over \partial
      W_\perp}\right)\Phi{\rm e}^{i(ky_t-\omega t)},
\end{equation}
where
\begin{equation}
  \label{eq:tp0}
  \Phi(z_t) \equiv \int_{-\infty}^t\hp(z(\tau)){\rm e}^{-i\omega'(\tau-t)}d\tau,
\end{equation}
which is independent of $t$, but not of the final position on the
orbit $z_t$.  All positions and velocities appearing here, and from now
on, are for the unperturbed orbit.

\subsection{Shift mode perturbed distribution}

Now specialize to the case when the eigenmode is the shift mode $\hp(z(\tau)) =
-\Delta{d\phi_0\over dz}$. Using
\begin{equation}
  \label{eq:dphi0dx}
  {d\phi_0\over dz}= -{m_e\over q_e} v_z {dv_z\over dz}= -{m_e\over q_e}{dv_z\over dt},
\end{equation}
we have
\begin{equation}
  \label{eq:phiushift1}
  \Phi=\Delta{m_e\over q_e} 
  \int_{-\infty}^t{dv_z\over d\tau}{\rm e}^{-i\omega'(\tau-t)}d\tau
  =\Delta{m_e\over q_e} 
  \left\{\left[v_z{\rm e}^{-i\omega'(\tau-t)}\right]_{-\infty}^{\tau=t}
    + \int_{-\infty}^tv_zi\omega'{\rm e}^{-i\omega'(\tau-t)}d\tau \right\}.
\end{equation}
For a passing particle it is better to write $v_\infty$ for the
velocity at $\tau\to \pm\infty$ on the orbit under consideration and then
${dv_z\over d\tau}= {d\over d\tau}(v_z-v_\infty)$ so that
\begin{equation}
  \label{eq:phiushift}
  \Phi  =\Delta{m_e\over q_e} 
  \left\{(v_z(t)-v_\infty)
    + \int_{-\infty}^t(v_z-v_\infty)i\omega'{\rm e}^{-i\omega'(\tau-t)}d\tau \right\}.
\end{equation}
The advantage of this form is that $(v_z-v_\infty)$ is zero outside
the hole so the integral has a finite domain. For trapped particles
this same form applies provided that $\omega'$ has a positive
imaginary part that ensures convergence and eliminates the lower limit
contribution.  Taking $v_\infty=0$ is more convenient for trapped
particles.  

In any case, the first (integrated) term $(v_z-v_\infty)$
expresses the effect of an essentially rigid velocity-shift of the
trapped distribution function, because it gives rise to a term in
$\tf$
\begin{equation}
  \label{eq:fshift}
  m_ei\omega'\Delta v_z{\partial f_0\over\partial W_\parallel} =
  -m_ev_{hole} v_z{\partial f_0\over\partial W_\parallel}
  =-v_{hole}{\partial f_0\over \partial v_z}, 
\end{equation}
where $v_{hole}=-i\omega'\Delta=\dot\Delta$ is the hole incremental
parallel velocity (observed in the $v_y$ frame). The term involving
$v_y{\partial f_0\over \partial W_\perp}={\partial f_0\over
  m_e \partial v_y}$
will integrate to zero $dv_y$.

A natural notation for the other term, because it has dimensions
of velocity, is to denote it $\omega'\tL$ with
\begin{equation}
  \label{eq:Ltilde}
  \tL(\omega')\equiv  \int_{-\infty}^t(v_z-v_\infty)i{\rm
    e}^{-i\omega'(\tau-t)}d\tau.
\end{equation}
Although $\omega'\tL$ is locally of higher order in the transit time,
it cannot be simply ignored at positions outside the hole, because all
the other terms in $\Phi$ there are zero. It in fact gives rise to
the net perturbation $\tf$ that crosses the outer boundary of the hole
region. In other words, it is the source of the jetting
perturbation. Besides which, when we are dealing with situations where
$\omega'$ times the transit time is not small, the $\tL$ term contains
all the resulting new behavior, and it is the thing that is hard to
calculate.

\subsection{Shift mode rigid velocity-shift contributes no force}

\bigskip It is helpful to eliminate $\phi_0$ from the shift mode
expressions by using the orbit equilibrium relation eq.\
(\ref{eq:dphi0dx}), allowing us to express spatial variations in terms
of $v_z$. It is also helpful for passing particles to transform
distribution $v_z$-integrals into $v_\infty$-integrals.  Then at
constant $z$ (and $\phi$)
\begin{equation}
  \label{eq:videntities}
  v_\infty dv_\infty= v_z dv_z;\quad {\rm so} 
  \qquad {\partial v_z\over \partial v_\infty} = {v_\infty\over v_z}.
\end{equation}
Moreover, Vlasov's equation means $f_{0}(v_z)=f_\infty(v_\infty)$.

The non-adiabatic density perturbation $\tn=\int \tf d^3v$ acquires a
contribution $\tn_v$ from the first term $(v_z-v_\infty)$ of eq.\
(\ref{eq:phiushift}). However, its contribution to total force (integrated
over the hole) is zero because
\begin{equation}
  \label{eq:n1v}
  \tn_v\equiv\int iq_e\omega {\partial f_0\over \partial W_\parallel}
  \Delta {m_e\over q_e}(v_z-v_\infty) d^3v={i\omega\Delta }\int   {\partial
    f_\infty\over \partial v_\infty}\left(1-{v_\infty\over v_z}\right)dv_\infty;
\end{equation}
and so 
\begin{eqnarray}
  \label{eq:ntildev}
  \tF_v\equiv\int -q_e{d\phi_0\over dz} \tn_v dz &=& m_ei\omega\Delta \int\int  v_z
  {dv_z\over dz}  {\partial
    f_\infty\over \partial v_\infty}\left(1-{v_\infty\over
      v_z}\right)dv_\infty dz\nonumber\\
    &=&  m_ei\omega\Delta \int {\partial
    f_\infty\over \partial v_\infty}\left[\int  v_z
  \left(1-{v_\infty\over v_z}\right)dv_z\right] dv_\infty=0.
\end{eqnarray}
The same annihilation will occur for any quantity that can be rendered
into the form of a spatial integral ${dv_z\over dz} dz\to dv_z$ (the
square bracket here, which is zero) when all the
$z$-dependence of the integrand is in $v_z$. This proof is equally
valid for trapped particles as for passing. Therefore for momentum
balance purposes we can ignore $\tF_v$ and $\tn_v$.

The second term of eq.\ (\ref{eq:phiushift}), i.e.\ $\omega'\tL$, is not of
this form and gives non-zero total force. From now on, we shall
consider only shift modes and use $\Phi$ to denote just $\Delta
(m_e/q_e)\omega' \tL$.

\subsection{Low frequency jetting}

Major simplification of $\tL$ occurs for passing-particles when
$\omega'$ is much smaller than the inverse transit time of the
particles through the hole, because then the
${\rm e}^{i\omega'(\tau-t)}$ term can be taken to be unity. We defer
integration $d^2v_\perp$ over the perpendicular velocities and regard
$\omega'$ as fixed for now. The resulting contribution\footnote{The
  notation ${dG\over d^2v_\perp}$ denotes a quantity that when
  integrated $d^2v_\perp$ gives $G$.} to the perturbed density from
the ${\partial f_0\over \partial W_\parallel}$ term is
\begin{eqnarray}
  \label{eq:ntl}
  {d\tn_{p}\over d^2v_\perp}&=& \int m_e i\omega'{\partial f_0\over \partial
    W_\parallel}\Delta i\omega' \int^z_{z_s}(v_{z'}-v_\infty) {dz'\over v_{z'}}
           dv_z\nonumber\\
&=&  -\omega'^2\Delta \int{\partial f_\infty\over \partial
    v_\infty} \int^z_{z_s}\left(1-{v_\infty\over v_{z'}}\right) dz'
           {dv_\infty\over v_z}.
\end{eqnarray}
Here the lower limit of the $z'$-integral $z_s$ is the start of the
orbit, which depends on the sign of $v$, which we will denote
$\sigma_v$.\footnote{Consequently, for negative $v_\infty$, $dz'$ is
  negative, but $dz'/v_z$ is positive, as is $dv_\infty$. Thus for
  opposite signs of $\partial f_\infty\over \partial v_\infty$, the
  density perturbations arising from opposite velocities are
  opposite. They are also on opposite sides of the hole; so when
  multiplied by the antisymmetric potential gradient, opposite
  velocities give jetting force in the same direction.} Contributions
come only from places with $v_z\not= v_\infty$ i.e. $\phi\not=0$, so
$|z_s|$ must exceed the extent of the hole but need not actually be
$\infty$. We denote the other end position of an entire spatial
integral as $z_f$ (finish of orbit).

The passing particle force is then
\begin{eqnarray}
  \label{eq:FL}
  {d\tF_{p}\over d^2v_\perp} 
  &=& 
    {\;\sigma_v\hskip-5pt}\int_{z_s}^{z_f} -q_e{d\phi_0\over dz} {
      d\tn_{p}\over d^2v_\perp} dz \nonumber\\
  &=&
      -m_e \omega'^2\Delta  {\;\sigma_v\hskip-5pt}\int_{z_s}^{z_f} v_z{dv_z\over dz} 
      \int{\partial f_\infty\over \partial v_\infty}     
      \int^z_{z_s}\left(1-{v_\infty\over v_{z'}}\right) dz'
      {dv_\infty\over v_z} dz    \nonumber\\
  &=&  -m_e \omega'^2\Delta\int{\partial f_\infty\over \partial v_\infty}
      {\;\sigma_v\hskip-5pt}\int_{z_s}^{z_f} {dv_z\over dz}
      \int^z_{z_s}\left(1-{v_\infty\over v_{z'}}\right) dz' dz dv_\infty.
\end{eqnarray}
Now we do the $dz$ integral by parts, making it
\begin{equation}
  \label{eq:zparts}
  v_\infty\int^{z_f}_{z_s}\left(1-{v_\infty\over v_{z'}}\right)
    dz'-
   \int_{z_s}^{z_f} v_z \left(1-{v_\infty\over v_{z}}\right) dz
   =-\int_{z_s}^{z_f} v_z \left(1-{v_\infty\over v_{z}}\right)^2 dz.
\end{equation}
Reversing again the order of integration, we obtain a $dv_\infty$
integral that can be done by parts
\begin{equation}
  \label{eq:fparts}
  \int-{\partial f_\infty\over \partial v_\infty} v_z
  \left(1-{v_\infty\over v_{z}}\right)^2 dv_\infty
  =2f_sv_{z0}+\int f_\infty {d\over dv_\infty}\left[  v_z
  \left(1-{v_\infty\over v_{z}}\right)^2 \right] dv_\infty,
\end{equation}
where $v_{z0}$ is the absolute value of $v_z$ at the separatrix, and $f_s$ is
the distribution function on the separatrix. The leading term arises
from the sign discontinuity of $v_{z}$ at $v_\infty=0$. Now
\begin{eqnarray}
  \label{eq:dbdvi}
  {d\over dv_\infty}\left[  v_z
  \left(1-{v_\infty\over v_{z}}\right)^2 \right] 
  &=&
      \left[{v_\infty\over v_z} \left(1-{v_\infty\over v_{z}}\right)
      -2\left(1-{v_\infty^2\over
      v_z^2}\right)\right]\left(1-{v_\infty\over v_{z}}\right)\nonumber\\
  &=& -2 +3{v_\infty\over v_{z}}- {v_\infty^3\over v_{z}^3}.
\end{eqnarray}
Hence
\begin{equation}
  \label{eq:FLfinal}
 {d\tF_{p}\over d^2v_\perp}=-m_e\omega'^2\Delta{\;\sigma_v\hskip-5pt}\int_{z_s}^{z_f} 2f_sv_{z0}+
\int \left[ -2 +3{v_\infty\over v_{z}}- \left(v_\infty\over v_{z}\right)^3\right]
  f_\infty dv_\infty dz,
\end{equation}
where ${\;\sigma_v}\int_{z_s}^{z_f}dz$ is simply $\int dz$.  This
agrees\footnote{Since $-\omega^2\Delta=(i\omega)^2\Delta=\ddot\Delta$
  which is the hole acceleration} with the prior kinematic
calculation \citep{Hutchinson2016}, except for the presence of the term
$2f_sv_{z0}$.

The trapped particle contribution associated with
${\partial f\over \partial W_\parallel}$ can be treated using the same
sequence of partial integrations except that the $v_\infty=0$ choice
means we cannot adopt $v_\infty$ as the velocity integration variable.
The past orbit integral now extends to $\tau=-\infty$, because the
orbit never escapes the hole, but assuming $\omega$ to have a positive
imaginary part, the integral converges and the lower limit can be
ignored\footnote{The orbit integral is written in eq.\ \ref{eq:FLt} in
  terms of $dz'=v_z' d\tau$. The start of the orbit $z_s$ is taken
  sufficiently far back \emph{in time} that the perturbing potential
  is negligible then, and since the $z'$ excursion in space is bounded
  for trapped orbits, we can ignore the lower limit.}. The phase-space element
${\partial f\over \partial W_\parallel} dW_\parallel={\partial
  f\over \partial v_z} dv_z$
commutes with the $dz$ integral. We interpret $dW_\parallel$ as
implying summing over positive and negative velocities, so we can
consider the $dz$ integral to be over the entire relevant orbit range
in the positive direction (avoiding the need for $\sigma_v$).
\begin{eqnarray}
  \label{eq:FLt}
  {d\tF_{t}\over d^2v_\perp}
  &=&
      m_e\Delta  \int v_z{dv_z\over dz} 
      \int i\omega'{\partial f_0\over \partial v_z}     
      i\omega' \int^z_{z_s}dz'
      {dv_z\over v_z} dz    \nonumber\\
  &=&  -m_e \omega'^2 \Delta\int{\partial f_0\over \partial v_z}
       \int {dv_z\over dz}
      \int^z_{z_s} dz' dz dv_z.\nonumber\\
  &=&   m_e \omega'^2 \Delta\int{\partial f_0\over \partial v_z}
      \int v_z  dz dv_z
      =m_e \omega'^2\Delta \int\int{\partial f_0\over \partial v_z}
       v_z dv_z dz \nonumber\\
  &=& m_e \omega'^2 \Delta \int
      \left[2f_sv_{z0}-\int_{-v_{z0}}^{v_{z0}}f_0 dv_z
      \right]dz
\end{eqnarray}
The term $2f_sv_{z0}$, which arises from the limits of
integration $dv_z$ at the separatrix, cancels the similar term in the passing
particle force expression (\ref{eq:FLfinal}). Without that term, the trapped force can be
considered to be simply the inertia of the trapped particles.

When $k=0$, $\omega'=\omega$, we can integrate $d^2v_\perp$, and no
${\partial f\over \partial W_\perp}$ term need be considered. Then we
find 
\begin{equation}
  \label{eq:jet1d}
  \tF =\tF_p+\tF_t= m_e \ddot\Delta  \int_{z_s}^{z_f} \left\{ \int\left[ -2 +3{v_\infty\over v_{z}}- \left(v_\infty\over v_{z}\right)^3
  \right]f_\infty dv_\infty+\int_{-v_{z0}}^{v_{z0}}f dv_z\right\}
 dz
\end{equation}
where $f$ here is the (unperturbed) one-dimensional
($v_z$)-distribution function.  This force expression is in full
agreement with the prior one-dimensional kinematic calculation \citep{Hutchinson2016}.

We observe that since an electron hole has no net electric charge, for
immobile ions the sum of the trapped electron charge and the
integrated difference of the passing electron charge density from its
external value must be zero. This allows us to deduce an alternative
expression for the trapped particle number:
\begin{equation}
  \label{eq:trapalt}
   \int_{z_s}^{z_f} \int_{-v_{z0}}^{v_{z0}}f dv_z dz=
    \int_{z_s}^{z_f} \int\left(1-\left|v_\infty\over v_z\right|\right)
    f_\infty dv_\infty dz
\end{equation}

\subsection{Low frequency momentum balance including  ${\partial f\over \partial W_\perp}$ terms}

Let us suppose that the external distribution is isotropic Maxwellian
(and the hole is stationary in the Maxwellian frame). Then the $kv_y$
terms in $\tf$ cancel each other for passing particles which has the
effect of making one of the $\omega'$ terms in eq.\ (\ref{eq:FLfinal})
just $\omega$.  No such cancellation occurs for the trapped particles
because the sign of $df/dW_\parallel$ is reversed: the electron
distribution is smaller at smaller $v_z$. For shallow holes for
trapped particles $|df/dW_\parallel| \gg |df/dW_\perp|$ so the
$df/dW_\perp$ term hardly contributes.  A simple way to account for it
is to suppose that the distribution is of the Schamel
type \citep{Schamel1979} having a parallel Maxwellian of negative
temperature in the trapped region. In that case,
\begin{equation}
  \label{eq:schameltemp}
  {\partial f\over \partial W_\parallel} = \beta   {\partial
  f\over \partial W_\perp},
\end{equation}
where $\beta$ is the inverse of the ratio (a negative quantity) of
trapped parallel temperature to perpendicular. This ansatz is very
convenient because it allows us simply to multiply
$kv_y{\partial f\over \partial W_\parallel}$ by $(1+{1\over|\beta|})$
to account for the $ {\partial f\over \partial W_\perp}$ term. This
extra factor accompanies all $k^2$ terms.  For a slow-moving hole,
$-\beta\approx 1+(15/16)\sqrt{\pi T_e/e\psi}$, which is large for a
shallow hole (small $\psi$). But the quantity $1/|\beta|\approx 0.346$
is still only a moderate correction for a deep hole $\psi=T_e/e$.

We still need to integrate over $v_y$ to arrive at the total force.
When we do so, first order $kv_y$ terms coming from the cross products
$\omega kv_y$ integrate to zero for a symmetric $f(v_y)$ distribution.
The passing particle force $\tF_p$ then is unchanged in form except
that only $\omega$ appears in it, not $\omega'$. The trapped force
$\tF_t$ has an $\omega^2$ term that adds to the passing as before,
plus a $k^2\langle v_y^2\rangle$ term that is otherwise of the same
form as eq.\ (\ref{eq:FLt}).

Consequently, using $\langle v_y^2\rangle=T_y/m_e$, and considering
$f$ to be the parallel distribution function, the full particle force
can be taken as 
\begin{eqnarray}
  \tF&=&-m_e \Delta  \int \omega^2\left\{ \int\left[ -2 +3{v_\infty\over v_{z}}- \left(v_\infty\over v_{z}\right)^3
         \right]f_\infty dv_\infty+\int_{-v_{z0}}^{v_{z0}}f
         dv_z\right\}\nonumber\\
     && \qquad\qquad+(1+{1\over|\beta|})k^2{T_y\over m_e} 
        \left\{-2v_sf_s+ \int_{-v_{z0}}^{v_{z0}} f
        dv_z\right\}\; dz.\\
  &=&-m_e \Delta  \int \omega^2\left\{ \int\left[ -1
      +2{v_\infty\over v_{z}}- \left(v_\infty\over
      v_{z}\right)^3\right]f_\infty dv_\infty\right\}\nonumber\\     
     && \qquad\qquad+(1+{1\over|\beta|})k^2{T_y\over m_e} 
        \left\{\int_{-v_{z0}}^{v_{z0}} (f-f_s)
        dv_z\right\}\; dz\nonumber\\
        &=& m_e \Delta n_\infty \int  \left[
            \omega^2h(\chi)+(1+{1\over|\beta|})k^2{T_y\over
            m_e}g(\chi)\right] dz 
  \label{eq:kinkforce}
\end{eqnarray}
The curly brace expressions have been denoted by the dimensionless
functions $h$ and $g$, which depend upon potential $\phi$ expressed in
terms of $\chi^2=-q_e\phi/T_e=m_ev_{z0}^2/2T_e$. They are both
positive.  In so far as the electric field stress is negligible, the
dispersion relation is $\tF=0$ which immediately shows that
$\omega^2= - (\langle g\rangle/\langle h\rangle) (1+{1\over|\beta|})k^2T_y/m_e$, where
$\langle g\rangle$ and $\langle h\rangle$ are the spatial averages of
$g$ and $h$. The frequency is therefore pure imaginary, one root being
positive, which is a growing unstable perturbation.

For an unshifted Maxwellian $f_\infty$, the velocity integrals needed for eq.\
(\ref{eq:kinkforce}) can be carried out. They are
\begin{equation}
  \label{eq:hofchi}
  h(\chi)=-2\int_0^\infty 
  \left[ -1 +2{v_\infty\over v_{z}}- \left(v_\infty\over v_{z}\right)^3
  \right]{f_\infty\over n_\infty} dv_\infty=-{2\over\sqrt{\pi}}\chi +(2\chi^2-1){\rm
    e}^{\chi^2}{\rm erfc}(\chi) +1
\end{equation}
and (using eq.\ \ref{eq:trapalt})
\begin{equation}
  \label{eq:gofchi}
  g(\chi)=2\left\{{f_s\over n_\infty}v_{z0}-\int_0^\infty 
    \left(1-\left.v_\infty\over v_z\right.\right)
    {f_\infty\over n_\infty} dv_\infty
  \right\}={2\over\sqrt{\pi}}\chi-
  \left[1-{\rm e}^{\chi^2}{\rm erfc}(\chi)\right]
\end{equation}
For small $\chi$ the power series is ${\rm erfc}(\chi) = 1-{2\over\sqrt{\pi}}
[\chi-\chi^3/3+O(\chi^5)]$, so $h(\chi)=\chi^2-{2\over\sqrt{\pi}}{4\over
  3}\chi^3+O(\chi^4)$ and $g(\chi)=\chi^2-{2\over\sqrt{\pi}}{2\over
  3}\chi^3 +O(\chi^4)$. The ratio $g/h\to 1$ as $\chi\to 0$; at $\chi=1$,
$g/h= 1.86$ and $\langle g\rangle/\langle h\rangle=1.63$ (for a ${\rm
  sech}^4z$ shape potential).

The electric tension force when  $\phi_0=\psi\,{\rm sech}^4(z/4\lambda_D)$
can be evaluated as
\begin{equation}
  \label{eq:etension}
  F_E= \Delta \epsilon_0 k^2 \int \left(d\phi_0\over dz\right)^2 dz=
\Delta \epsilon_0 k^2{\psi^2\over \lambda_D}{128\over 315}
= \Delta k^2\lambda_D {q_e^2\psi^2n_\infty\over T_e}{128\over 315}
\end{equation}
Therefore the full dispersion relation is
\begin{equation}
  \label{eq:fulldis}
  m_e\omega^2\int h(\chi)dz =  - k^2T_y \left((1+{1\over|\beta|})\int g(\chi) dz 
-\lambda_D {q_e^2\psi^2\over
    T_yT_e}{128\over 315}\right) 
\end{equation}
A crucial observation is that as $\psi\to 0$,
$g,h\to \chi^2\sim \psi$, whereas the electric tension scales like
$\psi^2$. Therefore for shallow holes ($\psi\ll eT_e$) the $F_E$ term
is ignorable. Even for a deep hole such as $\psi=eT_e$, numerical
evaluation shows that $\int g(\chi) dz= 0.55\times 4\lambda_D$, which
makes the $g$ term 5.4 times larger than $F_E$. Therefore for all but
exceptionally deep holes, $F_E$ can be ignored for low-$k$ modes.

In summary then, ignoring $F_E$, the predicted instability at low $k$,
where the transit, or bounce, time of the electrons is short compared
with $1/\omega'$, is that the imaginary part of the frequency is
\begin{equation}
  \label{eq:disperslow}
  \omega_i= k \sqrt{{T_y\over m_e}}\sqrt{{\langle g\rangle\over\langle h\rangle}(1+{1\over|\beta|})}
\end{equation}
where the second square root factor is unity for shallow holes and
rises only to 1.48 for $\psi\simeq T_e/e$.
(The real part of the frequency is zero.)

In view of the proportionality of $\omega_i$ and $k$, one expects that
the fastest growing mode has large $k$. However, the calculation so
far is for low-$k$. Therefore the observed growth rate in a
simulation or in nature is anticipated to be at the upper end of the
$k$-range for which the low-$k$ approximations apply. We therefore
need to analyze the breakdown of the low-$k$ approximations, to find the
behavior of the fastest growing modes.

\subsection{Full dispersion relation including finite transit time effects}

In order to determine the behavior at high-$k$ near the instability
threshold, we must consider situations where the transit time is
comparable to $1/\omega'$. We must therefore abandon the low-$k$
approximation and fully account for $\omega'$ in the integral $\tL$ in
eq.\ (\ref{eq:Ltilde}). Rather than pursue further analytical
approximation, numerical integration is adopted. Since this requires
a quadruple integration over $\tau$, $v_y$, $v_z$, and $z$, which
becomes computationally expensive if not done efficiently, it is
helpful to recognize that one can actually combine the evaluations
corresponding to all positions $z$ into a single orbit integral
$d\tau$. In other words, one does not have to do a \emph{different}
orbit integral for every position $z$ for a certain parallel energy
$v_z^2$. That integral can be done once, accumulating values for all
positions $z$ and can be scaled to provide $\tL$, $\Phi$ and hence
$\tf$ and the force contribution.

\subsubsection{Evaluation of the past orbit integral for passing particles}

For passing particles of specified $v_\infty$, I perform spatial
integrals on a uniform $z$-grid, starting at a negative $z$ position
$z_s$ far enough outside the hole to have its integrand negligible.
Call the past orbit time there $\tau_s=0$. For each succeeding
position $z_j=z_s+j\delta z$, I find its corresponding orbit time as
$\tau_{j+1}-\tau_j=\delta\tau=\delta z (1/v_j+1/v_{j+1})/2$ which is
an appropriate trapezoid increment for $\tau=\int dz/v$. The $v_j$
depend only on the known potential $\phi_j$ at $z_j$, and on
$v_\infty$. The next value of
$\tL\equiv\int_{\tau_s}^{t}(v-v_\infty)i{\rm e}^{-i\omega'(\tau-t)}d\tau$ is
calculated writing $v_{j+1/2}=(v_{j+1}+v_j)/2$ and using
\begin{eqnarray}
  \label{eq:Lstep}
  \tL_{j+1}-{\rm e}^{i\omega'\delta\tau}\tL_j&\approx&
  (v_{j+1/2}-v_\infty)\int_{\tau_j}^{\tau_{j+1}}
   i{\rm e}^{-i\omega'(\tau-\tau_{j+1})}d\tau\nonumber\\
  &=& 
-(v_{j+1/2}-v_\infty){1\over \omega'}
  \left[1-{\rm e}^{i\omega'\delta\tau}\right].
\end{eqnarray}
This approach preserves accuracy since $v$ is slowly varying even if
${\rm e}^{-i\omega'\tau}$ is not. Thus all values of $\tL_j$ and hence
of $\Phi$ (eq. \ref{eq:phiushift}) and the contribution to the hole
force on the entire uniform $z$-mesh are obtained from one cumulative
integral procedure, that gives
\begin{equation}
  \label{eq:dFpL}
  {dF_p\over d^2v_\perp dv_\infty} = 
  im_e\Delta\left(\omega'
    {\partial f_0\over\partial W_\parallel} +(\omega-\omega'){\partial f_0\over \partial
      W_\perp}\right)\int-q_e{d\phi\over dz}\omega' \tL(z,\omega') {dz\over v_z} v_\infty,  
\end{equation}
which can then be integrated over the (passing) velocity $v_\infty$ and
$v_y$ (and trivially $v_x$).

\subsubsection{Evaluation of the past orbit integral for trapped particles}
For trapped particles take $v_\infty=0$ and then denote orbits instead
by the quantity $v_\psi=\sqrt{2(-q_e\psi+W_\parallel)/m_e}>0$, which
is the orbit speed at $z=0$, $\phi=\psi$, when the parallel energy is
the negative quantity $W_\parallel(\ge q_e\psi)$. To obtain high
resolution near the separatrix, it is best to space orbits
by equal intervals of $\sqrt{-W_\parallel}$.
For a given $v_\psi$ the orbit has a finite $z$-extent, and turning
points ($v_z=0$) at its ends.  Equally-spaced $z$-positions are used,
but spanning just the $z$-extent of the orbit (different for each
energy $W_\parallel$). Near the ends of the orbit, integration
interpolation is optimized by determining $\delta\tau$ via
$\delta t= \delta v/\dot v $; but near $z=0$ using
$\delta t = \delta z/v$ is better.  Again $\tau$ integrals are
combined with $z$-integrals over the hole's spatial extent, by making
the $v_y$ integral the outermost and converting the combined
$dv_\psi dz$ phase-space integral of the orbit contribution over the
entire hole into an equivalent $d\tau$ integral round the closed
phase-space orbit.  Schematically
$F_t=\int\dots d\tau' dz dv_z dv_y= \int\dots d\tau' d\tau v_\psi
dv_\psi dv_y$.
\footnote{The integral over phase space between two orbits may be
  written $\int\int dzdv_z = \int\int d\tau v_\psi dv_\psi$, because
  by the incompressibility of phase space flow, the normal distance
  between adjacent orbits is $dv_\psi s_\psi/s$ where
  $s=\sqrt{v_z^2+(dv_z/dt)^2}$ is the speed of phase-space motion and
  $s_\psi=v_\psi$ (at the well center $\phi=\psi$ where $dv_z/dt=0$),
  so $dvdz\to sd\tau dv_\psi s_\psi/s = v_\psi d\tau dv_\psi $.}

As previously noted, one must assume that $\omega'$ has a positive
imaginary part $\omega_i$ which ensures the backward $\tau$ integral
giving $\tL$ converges. But it is obvious that the contribution from
each preceding orbit period is the same except that they are
multiplied by successive factors $\exp(i\omega't_b)$, where $t_b$ is
the orbit (bounce) period. This factor accounts for the attenuation
$\exp(\omega_i\tau)$ and the phase change relative to the succeeding
period. Therefore it is necessary to perform the integral numerically 
only around a \emph{single} period of the orbit; the total including
all the prior periods can then be synthesised by multiplying by the
infinite sum\footnote{This is a much more satisfactory numerical
  approach for the anharmonic orbits of a hole than expanding the
  motion as an infinite Fourier series. It retains the resonant
  behavior of the response, as can be seen by the fact that the
  denominator goes to zero where $\omega't_b=2\pi n$. See
   \citep{Lewis1982} for a formal discussion of this sort of expression.}:
\begin{equation}
  \label{eq:expsum}
  \sum_{\ell=0}^\infty [\exp(i\omega't_b)]^\ell = 1/[1-\exp(i\omega't_b)]
\end{equation}

It does not matter where one chooses to start and stop the single
orbit numerically evaluated. For definiteness, choose to
start the integral at $\tau=0$ saving the cumulative integral
$L(t)=\int_{0}^{t} v_z i{\rm e}^{-i\omega'(\tau-t)} d\tau$. Then
take $\tau=0$ to be the left end of the orbit (where $v_z$ changes
sign from negative to positive).  To construct the first prior orbit
for any other position $z,v_z$ on it, take $t$ to be the value of
$\tau$ during the first orbit at which the numerically calculated
orbit passes through that position.

Since the hole is reflectionally symmetric, it suffices to calculate
only half of the orbit, where $0<t\le t_b/2$. Starting at the left
hand end and integrating up to the right hand end is the same as
starting at the right and integrating along the negative-$v$ part of
the orbit to the left. The only difference is that the velocity is
opposite in sign. The resulting reversal of $\Phi$ is cancelled by the
reflection of the $z$ position and the consequent reversal of
$d\phi_0\over dz$. Hence positive and negative velocity parts of the
integral give equal contribution to the force.  The complete integral
for a full prior period ending in the $0<t\le t_b/2$ segment is
obtained as the sum of four parts of the orbit:
$0\to t\equiv t-t_b\to -t_b/2\to -t_b/2+t\to 0$. Noting that
$\int_{-t_b}^{t-t_b} v_z i{\rm e}^{-i\omega'(\tau-t)}d\tau={\rm
  e}^{i\omega't_b} L(t)$,
$\int_{-t_b/2}^{t-t_b/2} v_z i{\rm e}^{-i\omega'(\tau-t)}d\tau=-{\rm
  e}^{i\omega't_b/2} L(t)$
and
$\int_t^{t_b/2}v_z i{\rm e}^{-i\omega'(\tau-t)}d\tau= {\rm
  e}^{i\omega'(t-t_b/2)}L(t_b/2)-L(t)$ one finds
\begin{eqnarray}
  \label{eq:priorLR}
 \int_{t-t_b}^{t}{\hskip-10pt}
  &v_z(\tau)&{\hskip-10pt} i
              {\rm e}^{-i\omega'(\tau-t)}d\tau
              = \int_0^{t}+\int_{t-t_b}^{-t_b/2}
              +\int_{-t_b/2}^{-t_b/2+t}+\int_{-t_b/2+t}^0
v_z(\tau)i{\rm e}^{-i\omega'(\tau-t)}  d\tau
\nonumber\\
&=& 
 \quad L(t)  
  \qquad +\quad {\rm    e}^{i\omega't_b}[{\rm    e}^{i\omega'(t-t_b/2)}L(t_b/2)-L(t)]
\nonumber\\
&&- {\rm    e}^{i\omega't_b/2}L(t)
  - {\rm    e}^{i\omega't_b/2}[{\rm    e}^{i\omega'(t-t_b/2)}L(t_b/2)-L(t)]
\nonumber\\
&=&\left(1-{\rm e}^{i\omega't_b} \right) L(t)
    +{\rm    e}^{i\omega't}\left({\rm    e}^{i\omega't_b/2} -1 \right) L(t_b/2).
\end{eqnarray}
And the full prior time integral from $-\infty$ to give $\tL$ is obtained by
multiplying by eq.\ (\ref{eq:expsum}).

Then 
\begin{equation}
  \label{eq:dFtrapdvpsi}
  {dF_t\over d^2v_\perp dv_\psi} = 
  im_e\Delta\left(\omega'
    {\partial f_0\over\partial W_\parallel} +(\omega-\omega'){\partial f_0\over \partial
      W_\perp}\right)\int-q_e{d\phi\over dz}\omega' \tL(z,\omega') {dz\over v_z} v_\psi,
\end{equation}
where the integral $dz/v_z=dt$ is over the half-orbit with positive
$v_\psi$, and an equal contribution also comes from the negative-$v_\psi$ half.

\subsubsection{Results}

Implementing these algorithms gives the forces $F_t$ and $F_p$ for
specified general values of $\psi$, $k$, and $\omega$ for some
specified potential form $\phi_0(z)$ and distribution function $f$. A
verification of the numerics is obtained by comparing the
values found at $k=0$, $\Real(\omega)=0$, and $\Imag(\omega)$ small,
with the analytic expressions leading up to equation (\ref{eq:jet1d}),
and employing the forms (\ref{eq:hofchi}) and (\ref{eq:gofchi}).
\begin{figure} 
  \centering 
  \includegraphics[width=0.5\hsize]{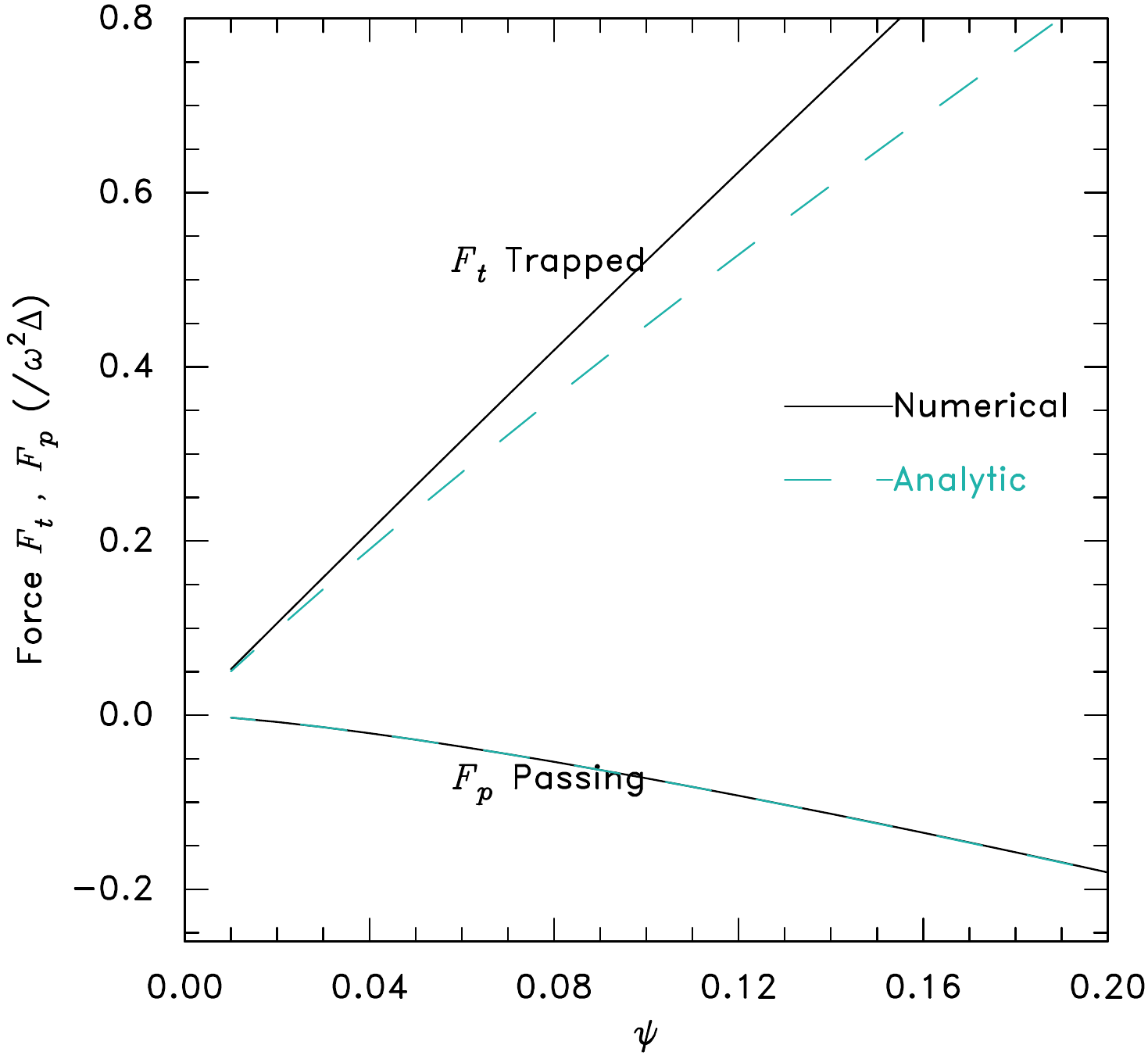}
  \caption{Comparison of numerical and analytic calculations of
    trapped and passing particle forces, for small value of $\omega_i$
    and zero $k$ and $\omega_r$. The forces are real for symmetric
    $v_y$-distributions.}
  \label{fig:verifyplot}
\end{figure}

As Fig.\ \ref{fig:verifyplot} indicates, I find essentially exact
agreement for the passing particle force $F_p$, using Maxwellian
distributions and a potential $\phi=\psi{\rm sech}^4(z/4)$ at all
values of $\psi$. However, the trapped particle force requires a
self-consistent parallel velocity distribution, which is not readily
available analytically except for shallow holes. Using the
approximation $f(W_\parallel)= f_\infty(0)\exp(-\beta W_\parallel)$ with
$-\beta= 1+(15/16)\sqrt{\pi T_e/e\psi}$, better than 2\% agreement in
$F_t$ is obtained for $\psi< 0.02$, which verifies the integration
coding.

\begin{figure} 
  \centering 
  \ (a)\hskip-20pt\includegraphics[width=0.45\hsize]{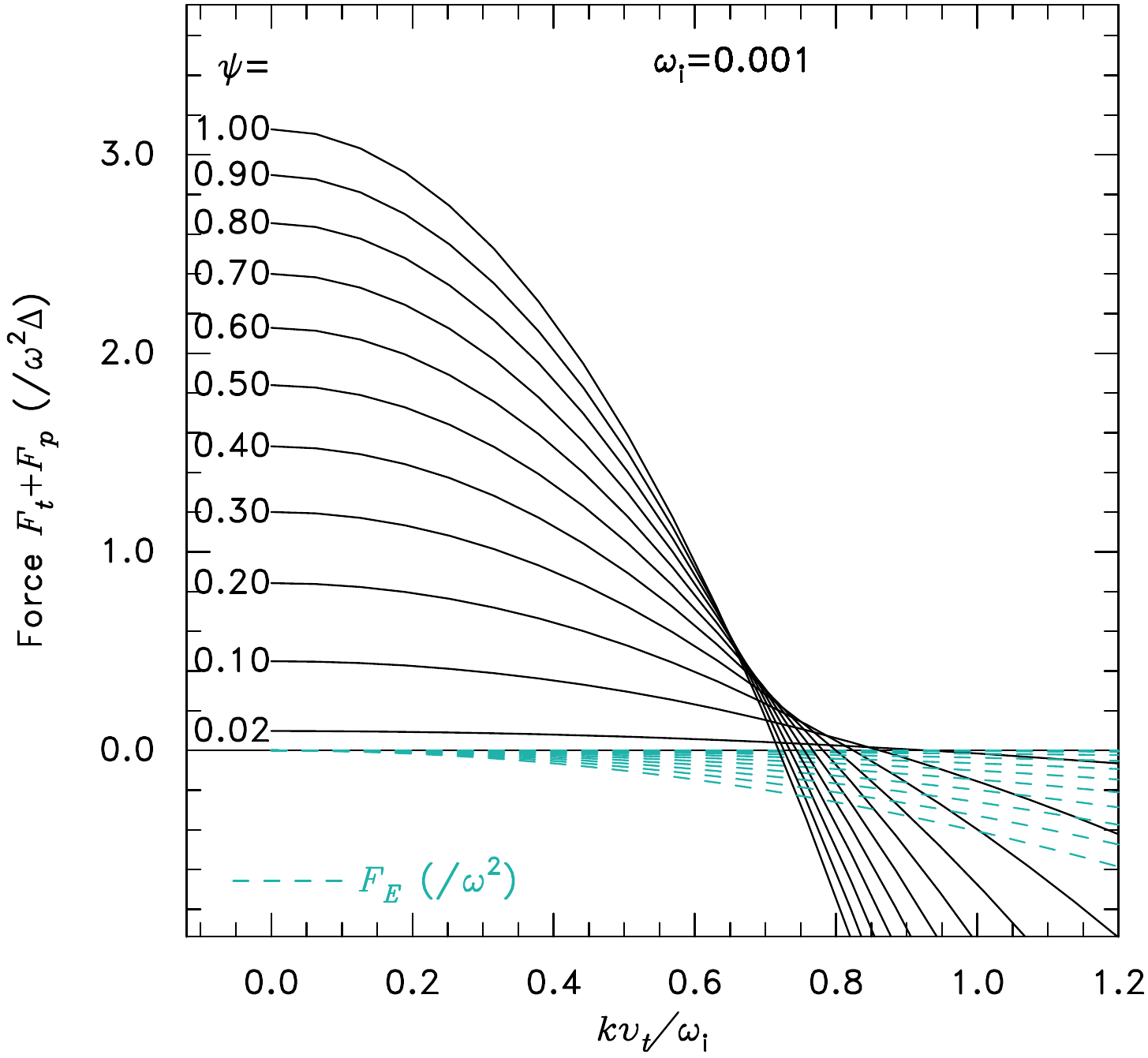}
  \ (b)\hskip-20pt\includegraphics[width=0.45\hsize]{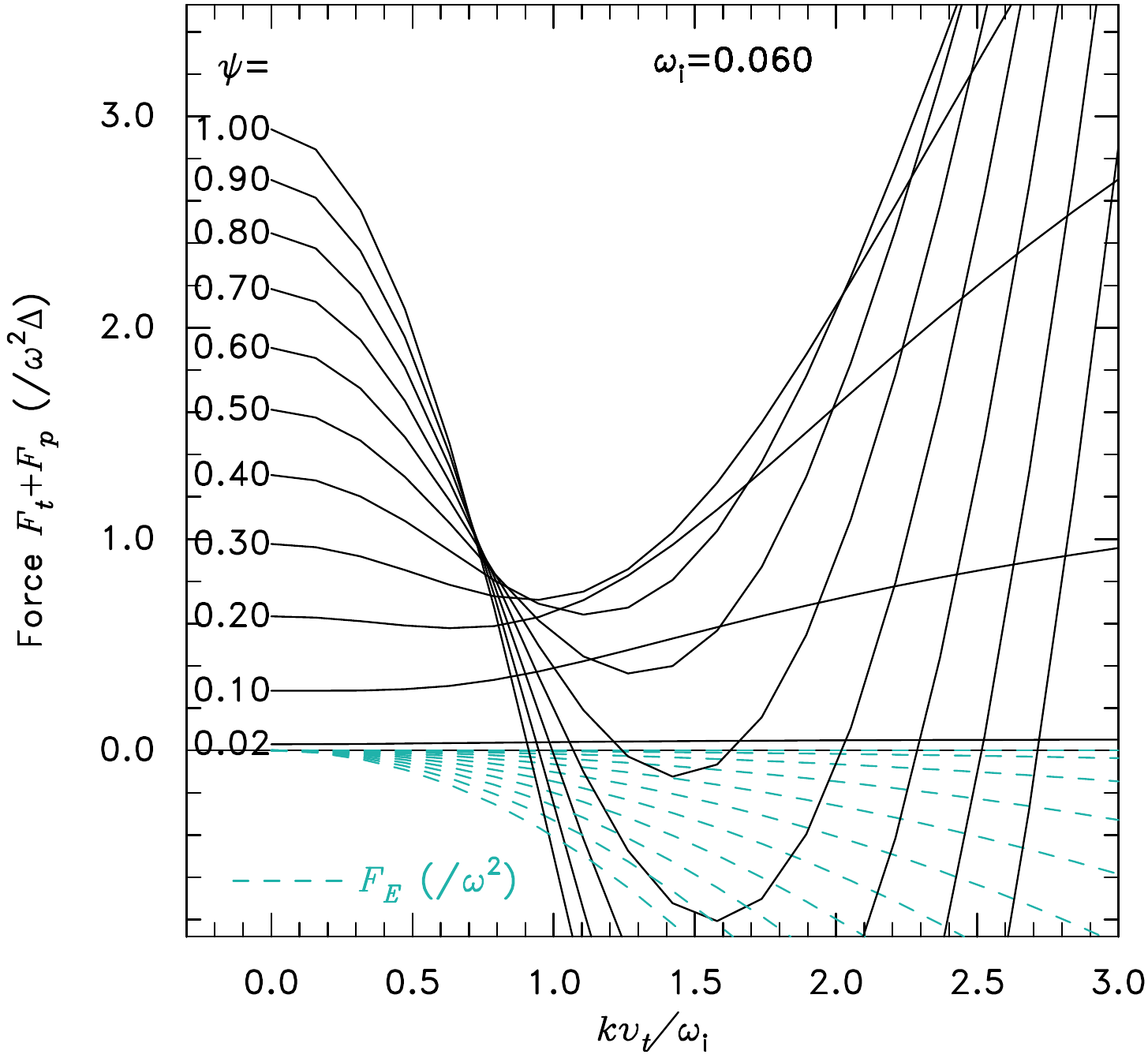}
  \caption{Total particle force $F_t+F_p$, and electric force $F_E$, normalized to $\omega^2$ versus
    wavenumber $k$. The dispersion relation is $F_t+F_p=F_E$.}
  \label{fig:Fvsk}
\end{figure}
As $k$ is increased from zero the main effect is observed to occur as
a reduction of $F_t$, with only small changes to $F_p$.  In Fig.\
\ref{fig:Fvsk} is shown the total particle force and the electric
tension force as a function of $k$ for two (fixed) imaginary
frequencies $\omega=i\omega_i$. It illustrates the fact that the
tension force $F_E$ (drawn for the corresponding $\psi$ values and
varying $\propto \psi^2$) is a minor correction, probably less
important in practice than the approximation arising from taking the
trapped distribution to be $\propto\exp(-\beta v_z^2)$. For small
$\omega_i$ (Fig.\ \ref{fig:Fvsk}(a)), this plot is essentially
universal. It gives the small-$k$ dispersion root at the intersection
of corresponding curves, where $kv_{t}/\omega_i$ is slightly
below 1 in accordance with eq.\ \ref{eq:disperslow}. However for
larger $\omega_i$ and correspondingly $kv_t$ (Fig.\
\ref{fig:Fvsk}(b)), there is no intersection for shallow holes (small
$\psi$) because the curves of $F_t+F_p$ have a minimum above
zero. Fig.\ \ref{fig:Fvsk}(b) also illustrates the existence of a
second solution at higher $k$ for higher $\psi$ values. We shall
shortly see that this behavior arises because of changes of resonance on
trapped electrons.

\begin{figure} 
  \centering
  \includegraphics[width=0.5\hsize]{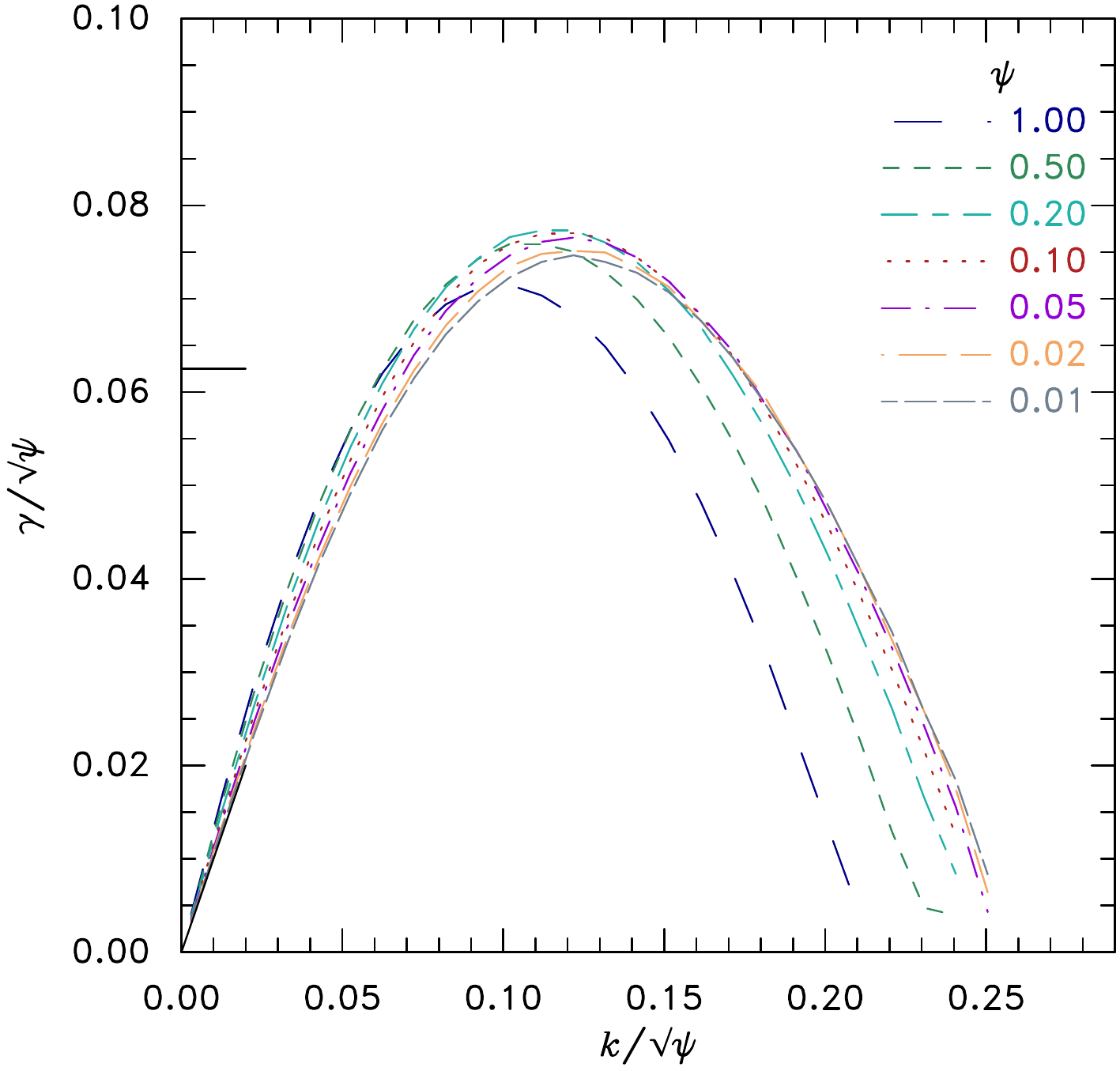}
  \caption{Dispersion relation showing scaled transverse instability
    growth rate versus scaled transverse wave number, for a wide range
    of hole potential $\psi$. Stationary hole,
    $\phi=\psi{\rm sech}^4(z/4)$, Maxwellian background, immobile
    ions. Units of $\gamma$, $z$ (and $k^{-1}$), and $\psi$ are
    respectively $\omega_p$, $\lambda_{De}$, and $T_e/e$.}
  \label{fig:gammavk}
\end{figure}
The dispersion relation that solves $F_t+F_p=F_E$ can usefully be
displayed in terms of $\gamma\equiv\omega_i$ as a function of given
$k$. This can be accomplished for a stationary hole in a Maxwellian
distribution by bisection real root-finding because the real part of
$\omega$ is zero by symmetry. Both $k$ and $\gamma$ scale like
$\sqrt{\psi}$, so plotting the resulting $\gamma/\sqrt{\psi}$ versus
$k/\sqrt{\psi}$, as shown in Fig.\ \ref{fig:gammavk}, gives an almost
universal curve, at least for $\psi\lesssim 0.5$.  [Here and
subsequently in presenting numerical results we use
non-dimensionalized parameters where units of time are $1/\omega_p$,
of space $\lambda_D$ and of energy $T_e/e$; that is equivalent to setting
$|q_e|=m_e=1$.] The iterative numerical solution breaks down at the
upper end of the $k$-range ($k_c$) where $\omega_i$ returns to zero,
because the integrals no longer converge. Numerically it breaks down
just before $k_c$, because of imprecision; that is why the curves do
not extend all the way back to the $k$ axis. The forces are evaluated
using the same first order approximation for $\beta$ which is
inaccurate at $\psi\sim 1$. That is one reason why the curves only
become truly invariant at low $\psi$. Another is that $F_E$ begins to
become significant for $\psi\sim 1$. Observe that the slope at the
origin also somewhat exceeds the limit value of 1 (indicated by the
short solid line) except when $\psi\ll 1$. That is consistent with
eq.\ (\ref{eq:disperslow}).  The peak growth rate occurs very close to
the previously published \citep{hutchinson18} estimate $k=\sqrt{\psi}/8$. The maximum
growth rate, even at low $\psi$ somewhat exceeds the prior estimate
$\gamma/\sqrt{\psi}=1/16$. This section has thus confirmed that the
transverse instability of electron holes (in Maxwellian background
plasma) is a kinematic effect arising from the jetting force-balance
in an accelerating hole; and it has given a detailed mathematical
derivation and precise numerical evaluation of coefficients.

\section{Magnetized Holes}

In the presence of a magnetic field (in the $z$-direction),
the unperturbed orbit is a helix. Because $v_z$ is not constant, the
helix has varying pitch, but since the wave vector is
perpendicular to $B$, that does not matter. Take the perpendicular
components of velocity at time $t$ to be $v_x(t)=v_\perp \cos\theta_t$
and $v_y(t)= v_\perp \sin\theta_t$, and write $\Omega=eB/m_e$
($q_e=-e$ so rotation is right-handed about $z$ for positive
$B$). Then at a different time $\tau$, the orbit is
\begin{eqnarray}
  \label{eq:orbit}
  v_x(\tau)&=& v_\perp \cos(\theta_t+\Omega[\tau-t])\quad,\quad
  v_y(\tau)= v_\perp \sin(\theta_t+\Omega[\tau-t]),\nonumber\\
  x(\tau) &=& x_t+(v_\perp/\Omega)
  \{\sin(\theta_t+\Omega[\tau-t])-\sin\theta_t\},\nonumber\\
  y(\tau) &=& y_t+(v_\perp/\Omega)
      \{-\cos(\theta_t+\Omega[\tau-t])+\cos\theta_t\}.
\end{eqnarray}

We substitute the orbit parameters into the
integral   (\ref{eq:f1energy}) and
evaluate it. The exponential's argument becomes
\begin{equation}
  \label{eq:exponential}
  i\left\{ky_t + \xi(\cos\theta_t-\cos[\theta_t+\Omega(\tau-t)])-\omega\tau
  \right\},
\end{equation}
where $\xi=kv_\perp/\Omega$. We can then write the $\tau$-dependence
using the Fourier expansion
\begin{equation}
  \label{eq:Bessexp}  
  {\rm e}^{-i\xi\cos\theta_\tau}=\sum_{m=-\infty}^\infty {\rm e}^{-im\theta_\tau}(-i)^mJ_m(\xi),
\end{equation}
(and its derivative with respect to the variable $\theta_\tau=
\theta_t+\Omega(\tau-t)$). From which we find
\begin{equation}
  \label{eq:Fourier}
  {\rm e}^{i(ky_\tau-\omega \tau)} = \sum_m (-i)^mJ_m(\xi)
  {\rm e}^{-i(m\theta_t -\xi\cos\theta_t)}
  {\rm e}^{i(ky_t+m\Omega t)}{\rm e}^{-i(m\Omega+\omega)\tau}
\end{equation}
and 
\begin{equation}
  \label{eq:Fourierprime}
  (kv_\perp/\Omega)\sin\theta_\tau{\rm e}^{i(ky_\tau-\omega \tau)} = -\sum_m m(-i)^mJ_m(\xi)
  {\rm e}^{-i(m\theta_t -\xi\cos\theta_t)}
  {\rm e}^{i(ky_t+m\Omega t)}{\rm e}^{-i(m\Omega+\omega)\tau}.
\end{equation}
Everything can be taken outside the $\tau$-integral for $\tf$, except
the final exponential and the quantity $\hp(z(\tau))$. So we define
the following quantity, which is independent of $t$ (but not $z(t)$),
\begin{equation}
  \label{eq:phim}
  \Phi_m(z)\equiv 
  \int_{-\infty}^t \hp(z(\tau)){\rm e}^{-i(m\Omega+\omega)(\tau-t)}d\tau,
\end{equation}
where $z(\tau)=z(t)+\int_t^\tau v_z(t')dt'$. The quantity $\Phi_m$ is
a partial-domain Fourier transform of $\hp(z(\tau))$ corresponding to
cyclotron harmonic $m$. It also has exactly the same form as the
unmagnetized $\Phi(z)$ with the identification
$\omega'=\omega+m\Omega$ (instead of $\omega'=\omega-kv_y$). The
remaining space and time $t$-dependence of $\tf$ (eq.\
\ref{eq:f1parts}) is then contained in a term
${\rm e}^{i(ky_t+\omega t)}$. The dependence on the gyrophase angle at
time $t$, $\theta_t$, is specified by the velocity $\v$ in the
argument of $f_1(\v,\x,t)$. The quantities we need to evaluate, such
as the density and force, require us to integrate over all $\v$, which
we can begin to do by integrating over $\theta_t$. In view of the
Bessel function Fourier expansions, we can write
\begin{equation}
  \label{eq:thetaexp}
  {\rm e}^{-i(m\theta_t -\xi\cos\theta_t)}=\sum_{\ell=-\infty}^\infty
  i^\ell{\rm e}^{i(\ell-m)\theta_t}J_\ell(\xi)
\end{equation}
which when integrated over $\theta_t$ gives $2\pi i^m J_\ell(\xi)$
when $\ell=m$, but zero otherwise. (Everything else in the $\tf$
integral is independent of $\theta_t$.) Consequently (dropping the $t$
suffix as indication of the final orbit position now, so we can use it
instead to denote thermal)
\begin{eqnarray}
  \int \tf d\theta/2\pi = i\sum_m \left[(m\Omega+\omega)
  {\partial f_0\over \partial W_\parallel}-m\Omega {\partial
  f_0\over \partial W_\perp}\right]q_e\Phi_m J_m^2(\xi)
 {\rm e}^{i(ky+\omega t)}
\end{eqnarray}


If the perpendicular distribution is Maxwellian
$f_{\perp0}=(m_e/2\pi T_\perp)\exp(-W_\perp/T_\perp)$, then the
integral with respect to $v_\perp$ is analytic
$\int_0^\infty J_m^2(\xi)v_\perp f_{\perp0}
dv_\perp=\exp(-\xi_t^2)I_m(\xi_t^2)/2\pi$,
where $\xi_t^2=k^2T_\perp/\Omega^2m_e$, and $I_m$ is the modified
Bessel function. Also
${\partial f_{\perp0}\over \partial W_\perp}=-f_{\perp0}/T_\perp$. Then there
results an expression for the perturbed \emph{parallel} distribution function
\begin{eqnarray}\label{eq:f1magnetic}
  f_{\parallel 1}(y,t) =  
  q_e\phi_1(t)\left.{\partial f_{\parallel0}\over\partial W_\parallel}\right|_t
  + \sum_m i\left[(m\Omega+\omega)
  {\partial f_{\parallel0}\over \partial W_\parallel}
  +m\Omega {f_{\parallel0}\over T_\perp}\right]
  q_e\Phi_m {\rm e}^{-\xi_t^2}I_m(\xi_t^2)
  {\rm e}^{i(ky+\omega t)}.
\end{eqnarray}
We naturally denote the second term (the sum) as
$\tf_\parallel {\rm e}^{i(ky-\omega t)}$.

\begin{figure} 
  \centering
  \includegraphics[width=0.5\hsize]{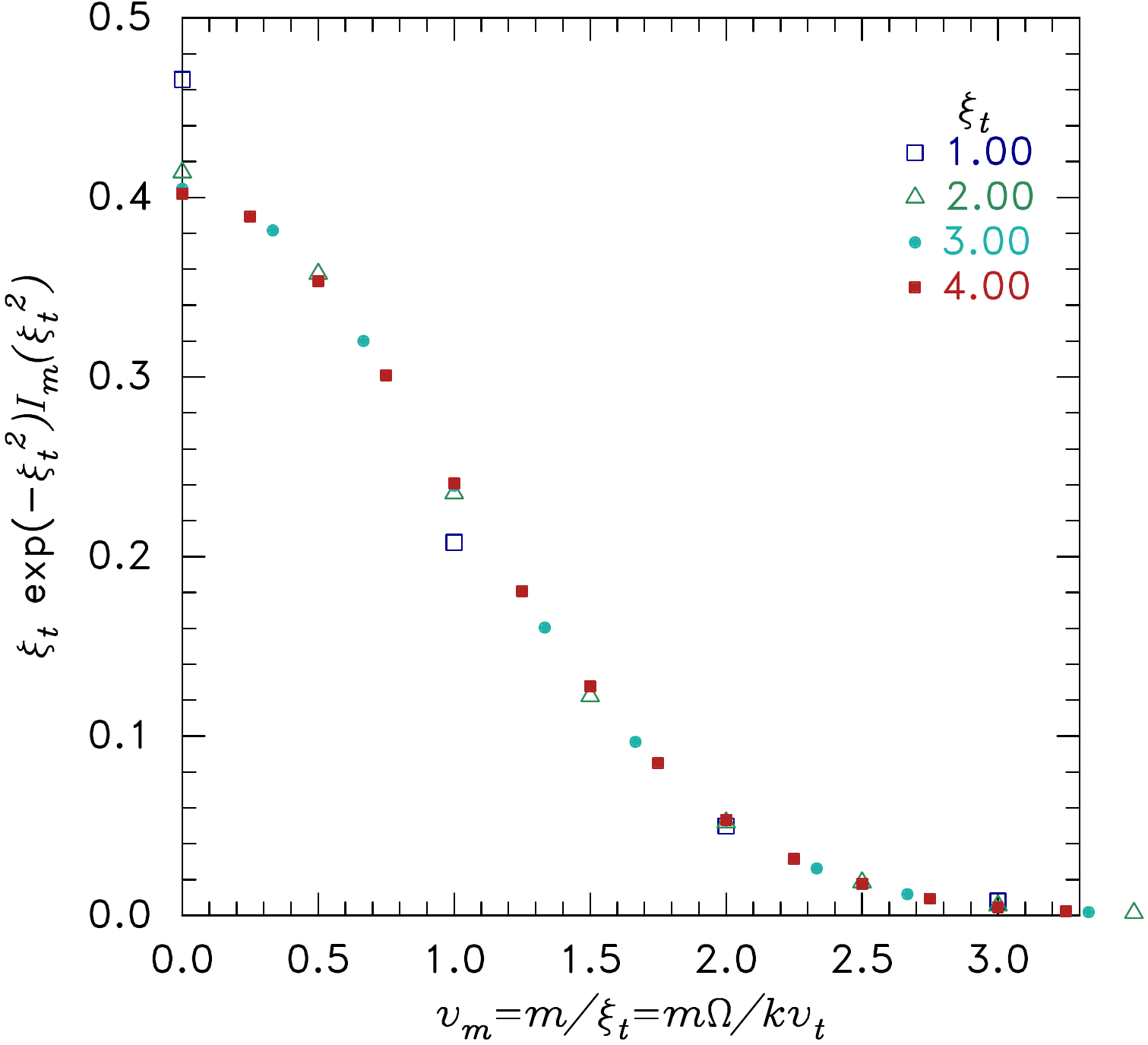}
  \caption{Illustration of the Bessel function harmonic sum
    approximating a transverse Maxwellian
    distribution.}
  \label{fig:bessosum}
\end{figure}

With this expression we have replaced the integral over perpendicular
velocities that was necessary for the unmagnetized case with a sum
over $\Phi_m{\rm e}^{-\xi_t^2} I_m(\xi_t)$. Therefore the numerical
effort involved in evaluating the force is no greater than what is
needed to evaluate the zero field case by numerical integration over
the perpendicular distribution (although it is limited to Gaussian
perpendicular distributions). Moreover, $\Phi_m$ is essentially
exactly the same quantity as the unmagnetized $\Phi$. Regarding
$\Phi(z,\omega',W_\parallel)$ as a function of $z$ and $\omega'$,
straightforwardly $\Phi_m(z)=\Phi(z,\omega+m\Omega,W_\parallel)$; with
$\omega'\to \omega+m\Omega$ (instead of $\omega-kv_y$). The close
connection to the unmagnetized calculation can be understood by
plotting $\xi_t {\rm e}^{-\xi_t^2}I_m(\xi_t^2)$ versus the transverse
phase velocity of harmonic $m$ as shown in Fig.\ \ref{fig:bessosum}.
Harmonics are spaced by a velocity increment (in units of the thermal
velocity $v_t=\sqrt{T_\perp/m_e}$) that is $\delta v=1/\xi_t=\Omega/kv_t$. Consequently the sum
$\sum_m{\rm e}^{-\xi_t^2}I_m(\xi_t^2) =\sum_m\xi_t {\rm
  e}^{-\xi_t^2}I_m(\xi_t^2) \delta v $,
when $\xi_t>1$, represents a kind of finite-difference approximation to
the integral $\int\xi_t {\rm e}^{-\xi_t^2}I_m(\xi_t^2) dv_m$, where
$v_m=m/\xi_t$ is regarded as a continuous variable for given $\xi_t$.
Fig.\ \ref{fig:bessosum} shows how quickly the values of the harmonic
sum approximates the integral over a Gaussian perpendicular
velocity distribution as the magnetic field becomes smaller. 

At low values of $\xi_t(<1)$ (high magnetic field), only one or two
harmonics give any substantial contribution.  In the infinite-field
limit, when only the $m=0$ harmonic matters (and $I_0(0)=1$), we
return to the one-dimensional motion expression:
\begin{equation}\label{eq:fp0}
  f_{\parallel 1}(t) =  
  q_e\phi_1(t)\left.{\partial f_{\parallel0}\over\partial W_\parallel}\right|_t
  + i\omega
  {\partial f_{\parallel0}\over \partial W_\parallel}
  q_e\Phi_0 {\rm e}^{i(ky_t-\omega t)}.
\end{equation}
which amounts to setting effectively $k=0$.  As $\Omega$ is decreased
from infinity ($\xi_t$ raised from zero), the $m=\pm 1$ terms next
become important. And for $\xi_t\ll 1$, since
${\rm e}^{-\xi_t^2}I_m(\xi_t^2) \simeq (\xi_t^2/2)^m/m!$, the
higher-$m$ terms are negligible (although up to at least $m=4$ are kept in the
numerics to retain precision). Noting
$\Phi(\omega')= (m_e\Delta/q_e)\omega'\tL(\omega')$, in eq.\
\ref{eq:f1magnetic} the $m=1$ relevant coefficient for small $\omega$
is
$m\Omega \Phi_m{\rm e}^{-\xi_t^2}I_m(\xi_t^2)=(m_e\Delta/q_e)\tL
\Omega^2 \xi_t^2/2 =(m_e\Delta/q_e) \tL k^2v_t^2/2$,
whose only dependence on $\Omega$ enters through $\tL(\omega')$, that
is through finite transit time effects. The particle force becomes
\begin{eqnarray}
  \label{eq:m0and1}
  \tF &=&
          -im_e\Delta \int q_e{d\phi_0\over dz} \left[ {\partial f_{\parallel
          0}\over\partial W_\parallel}\omega
          \sum_{m=0,\pm1}[\Phi_m {\rm e}^{-\xi_t^2}I_m(\xi_t^2)]\right.
          \nonumber\\ &&
      \qquad+\left.\left({\partial f_{\parallel0}\over\partial W_\parallel}
          +{f_{\parallel0}\over T_\perp}\right)
          \sum_{m=\pm1}m\Omega[\Phi_m {\rm e}^{-\xi_t^2}I_m(\xi_t^2)]\right] 
          dv_zdz\\
      &\simeq&-m_e\Delta\int q_e{d\phi_0\over dz} 
               \left[{\partial f_{\parallel0}\over\partial W_\parallel}
               \omega^2i\tL(\omega)
               +{\partial f_{\parallel t}\over\partial W_\parallel}
               \left(1+{1\over|\beta|}\right){1\over
               2}k^2v_t^2\sum_{\pm1}i\tL(m\Omega+\omega)      
               \right]dv_zdz,\nonumber
\end{eqnarray}
where terms of order $\omega/\Omega$ and higher are neglected,
${\partial f_{\parallel t}\over\partial W_\parallel}$ denotes the
trapped distribution's slope, and the passing $m=\pm1$ contribution
cancels by isotropy. This is identical to eq.\ (\ref{eq:kinkforce})
except insofar as the argument of $\tL$ is changed. Since $\tL$ is
constant in the short transit time limit, instability there is
\emph{not} stabilized by magnetic field. The low-$k$ instability
growth rate is still positive and given by (\ref{eq:disperslow})
unless $\Omega$ is at least of order the predominant bounce frequency
$\omega_b$, so as to change $\tL(\Omega+\omega)$. For definiteness we
take $\omega_b=\sqrt{\psi}/2$, which is the (dimensionless) bounce
frequency in a $\psi\,{\rm sech}^4z/4$ hole for trapped particles at
the bottom of the potential energy well.

What is required for stabilization is that (for small
$\omega=i\omega_i$) the value (which is real) of
$\sum_{m=\pm1}\int i\tL(m\Omega+i\omega_i){d\phi_0\over dz}dz
{df_{\parallel t}\over dW_\parallel} dv_z$,
for trapped particles be reversed, or else reduced sufficiently that
the combination with the electric field tension is reversed, so that
the dispersion relation $\tF_p=-\tF_t+F_E$ cannot be satisfied.

\begin{figure}
  \centering
  \includegraphics[width=0.5\hsize]{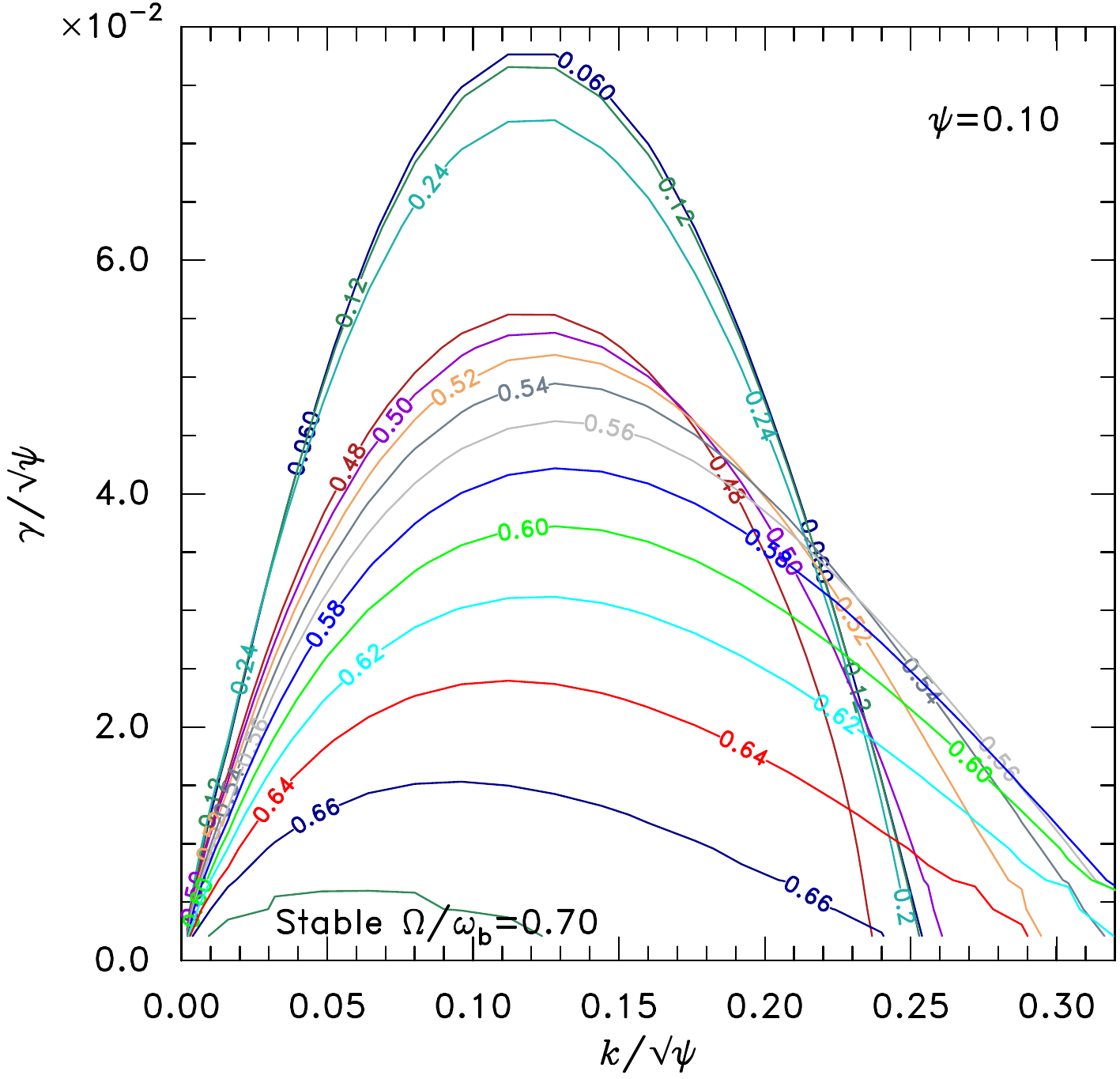}
  \caption{The zero contours of $F_p+F_t-F_E$ for magnetized holes labeled
    with the increasing values of $\Omega/\omega_b$. The contours give
    the the dispersion relation $\gamma(k)$. Magnetic stabilization of
    the transverse instability occurs as $\Omega/\omega_b\simeq 0.7$ is
    approached.}
  \label{fig:magstable}
\end{figure}

The total consequence of magnetization can be conveniently documented by
creating a contour plot of the total force $\tF_t+\tF_p-F_E$ over the
relevant domain of $\omega_i=\gamma$ and $k$. The sum of cyclotron harmonics
replaces the $\int dv_y$ and is used to evaluate the force for the
entire hole. Only the zero-contour of a series of contour plots for
different $\Omega$ is shown in Fig.\ \ref{fig:magstable}, since zero
value occurs at the solution of the dispersion relation (when there is
one). The labels are the values of $\Omega/\omega_b$.
Once again, this is a nearly universal figure for the
${\rm sech}^4(z/4)$ hole, at least at low $\psi$ where the $\beta$
approximation for the trapped energy distribution is accurate.

\begin{figure} 
  \centering
  \noindent (a)\hskip-18pt\includegraphics[width=0.45\hsize]{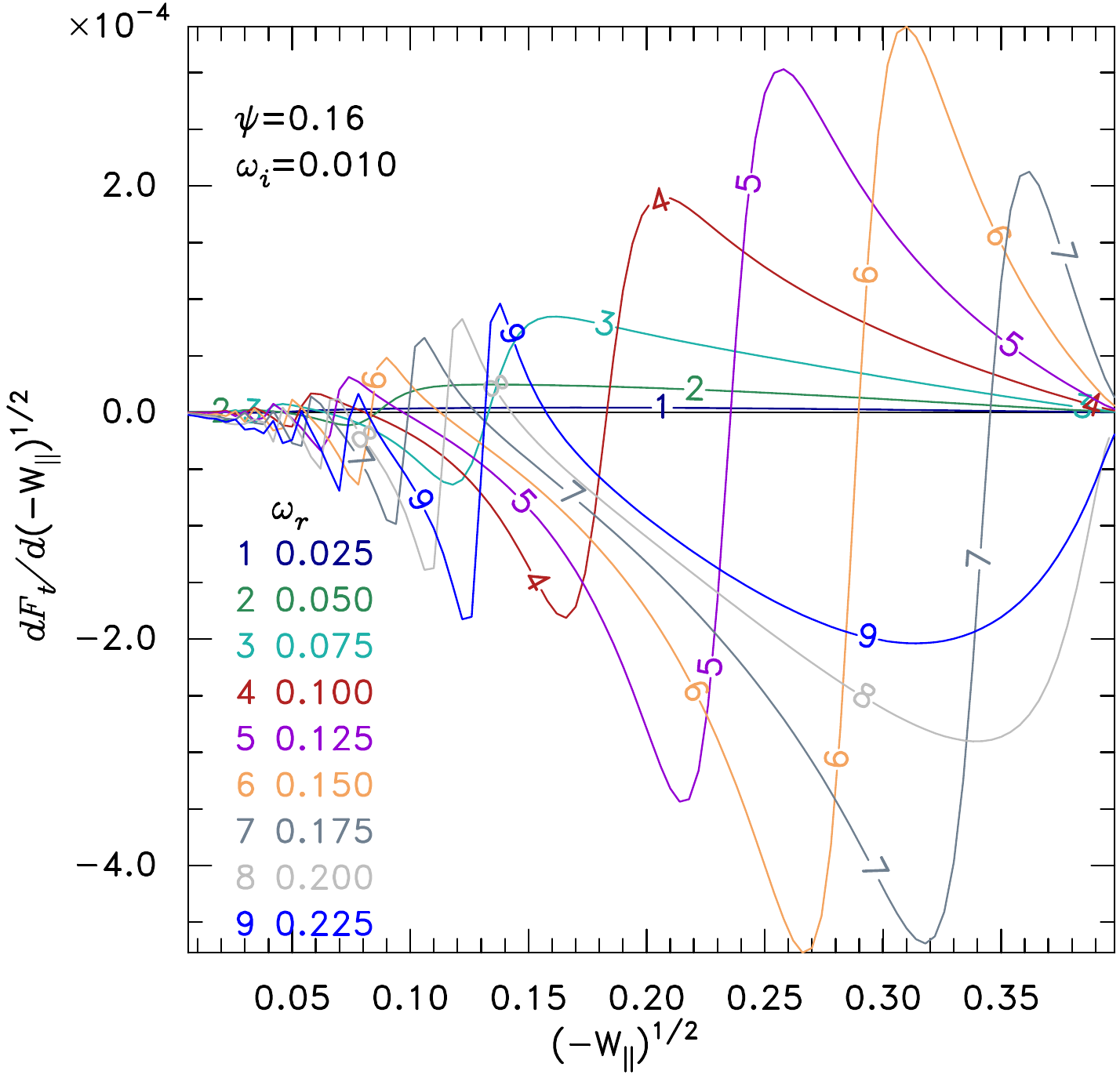}\ 
  \noindent (b)\hskip-18pt\includegraphics[width=0.45\hsize]{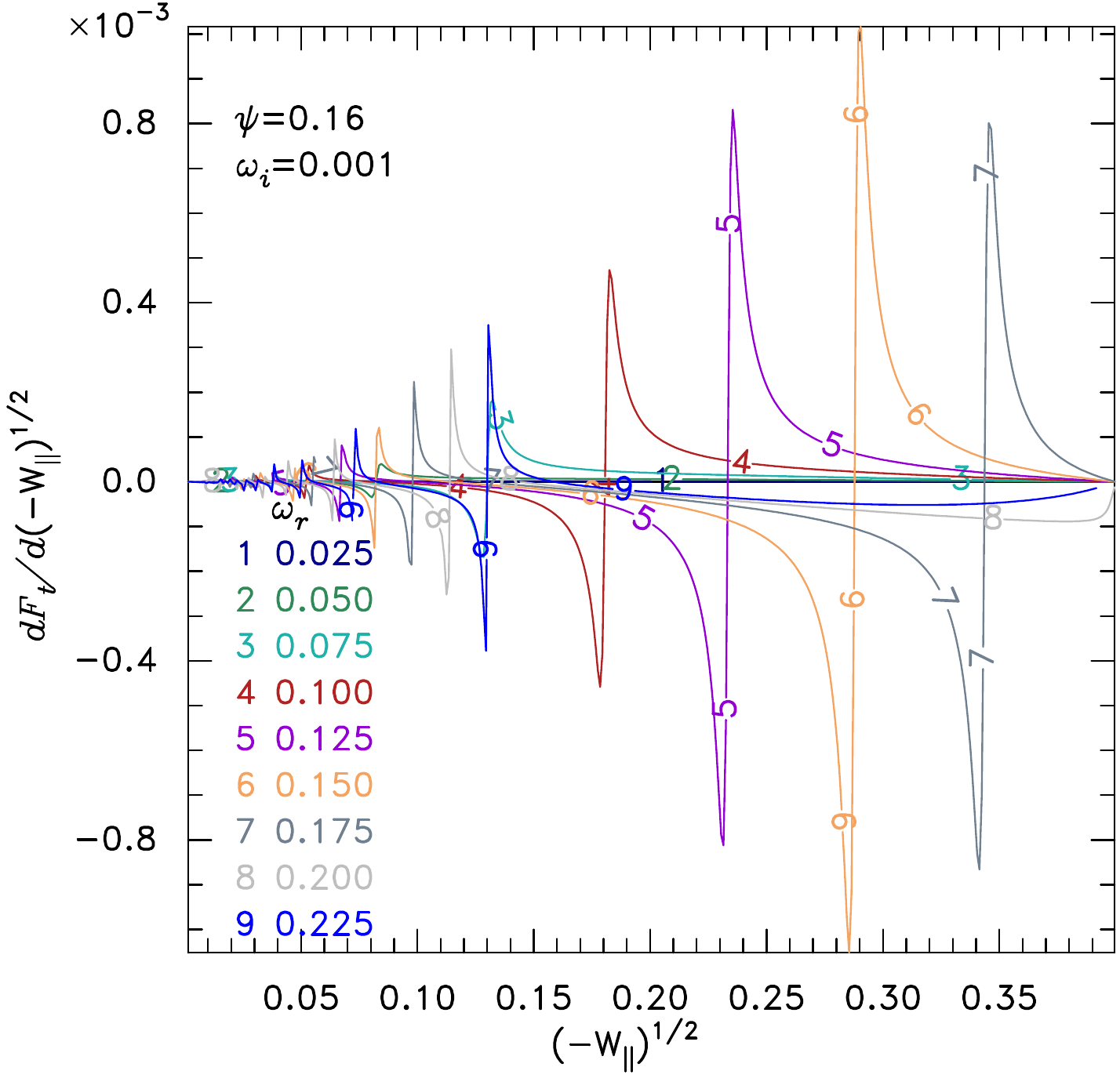}
  \caption{Contributions to the trapped particle force for specific
    real and imaginary parts of $\omega'$, as a function of trapped
    energy. The real part $\omega_r$ corresponds to $\Omega$. The effects of
    bounce resonances dominate the variation at higher
    $\omega_r$. Resonances are sharper at lower $\omega_i$, (b), than at
    higher, (a).}
  \label{fig:dFt}
\end{figure}

Stabilization arises from a reversal of $\tF_t$, which can be
understood by examining the contributions to the trapped force from
different trapped energy particles. Fig.\ \ref{fig:dFt} shows how the
contribution to trapped force from different parallel energy orbits
varies for different real parts ($\omega_r$) of $\omega'$ in eq.\
(\ref{eq:dFtrapdvpsi}). It is expressed as $dF_t/d(-W_\parallel)^{1/2}$ versus
$(-W_\parallel)^{1/2}$, so that the total force is the area under the
curve, integrated from zero to $\sqrt{\psi}$ (the right hand end).  
The application to a magnetized case is that the cyclotron harmonic
trapped force contribution is equal to the same integral but with
$\omega_r=\Omega$. So this plot represents how the ($m=\pm1$) trapped
force depends on parallel energy. The behavior is dominated by
resonances between the frequency $\omega_r$ and the bounce frequency
of the particles (and its harmonics, since the bounce
orbits are anharmonic). The bounce frequency varies from zero for
marginally trapped particles ($W_\parallel=0$) to the value
$\omega_b=\sqrt{\psi}/2$ for the deepest trapped particles
$W_\parallel=\psi$. The resonant energy corresponds to the place where
the curve passes rapidly through zero. For the unmagnetized case,
contributions for $kv_y=\omega_r$ are represented by the same plot.

Low $\omega_r$ is resonant only at very low bounce frequency, that means low
$|W_\parallel|$: marginally trapped particles at the left hand end of
the plot. Thus the areas under curves 1 and 2 in Fig.\
\ref{fig:dFt}(a) are dominated by regions to the right of the
resonances, where $dF_t/d(-W)^{1/2}$ is positive. Curves 3 and 4 show
the fundamental resonance approaching the center of the energy range;
so significant contribution is beginning to arise from regions to the
left of the fundamental resonance where $dF_t/d(-W)^{1/2}$ is
negative. Curves 5 and 6 show the transition to the state where the
negative contributions exceed the positive, and curve 8 corresponds to
$\omega_r = \sqrt{\psi}/2$ where the resonance is exactly at the right
hand end of the energy range: particles at the bottom of the potential
well. It has a clearly predominant negative integrated force. For the
higher frequencies, e.g. 8 and 9, it is easy to see the second (and
higher) bounce-frequency harmonic resonance effects in the left-hand one
third of the energy range.

All these qualitative statements depend little on the choice of other
parameters. The hole depth $\psi$ determines the energy range of the
plot; the imaginary part of $\omega'$, $\omega_i$, determines the
sharpness of the resonance, as is seen by comparing Fig.\
\ref{fig:dFt}(a) with \ref{fig:dFt}(b). Low $\omega_i$ requires fine
energy mesh to resolve accurately. Once $\omega_r>\omega_b$ the first
bounce harmonic resonance disappears but stability remains, showing
that cyclotron damping \emph{per se} is not the stabilization
mechanism, since no particles are then resonant.

\section{Summary}

The transverse instability of an electron hole in an isotropic
Maxwellian background plasma arises not because of orbit transverse
`focusing' but because of the overall hole force balance arising from the
kinematic parallel jetting (energization) of particles. The shift mode
is the relevant low frequency perturbation eigenmode, proportional to
the spatial derivative of the hole potential. It is demonstrated that
a linearized Vlasov-Poisson calculation of the jetting force gives
expressions identical at low frequency to those recently derived by
more elementary analysis.  Previous stability analyses of
\emph{symmetric} eigenmodes are irrelevant to the long-wavelength
shift mode which kinks the hole, because the shift-mode is
\emph{anti-symmetric}.  A rigorous linear analysis of the dispersion
relation of the shift mode has been completed for a slowly moving
electron hole, ignoring ion motion. The instability is purely growing
and $\gamma(k)$ has been found and expressed in essentially universal
form. The peak growth rate occurs, for a
$\psi\, {\rm sech}^4(z/4\lambda_D)$ hole potential, at
$k\lambda_D\simeq \sqrt{e\psi/T_e}/8$, with a value (in the absence of
magnetic field) slightly above $\gamma=\omega_p \sqrt{e\psi/T_e}/16$,
in good confirmation of the recent estimates \citep{hutchinson18} based on the same
kinematic mechanism. The magnetized case has also been analysed,
showing that stabilization occurs at $\Omega\simeq 0.7 \omega_b$ where
$\omega_b=\omega_p\sqrt{e\psi/T_e}/2$ is the bounce frequency of
deeply trapped electrons. Stabilization occurs because the jetting
force on the trapped electrons reverses its sign across the resonance
between the bounce and cyclotron frequencies.  The same sort of
reversal effect is responsible for suppressing the unmagnetized
instability when $kv_t \gtrsim\omega_b$. This upper $k$-limit suggests
that holes with tranverse extent less than approximately
$4\pi \lambda_D\sqrt{T_e/e\psi}$ should not suffer this instability.

There remain some discrepancies between the present analysis and
simulation observations. In Maxwellian plasmas it has been documented
in \citep{hutchinson18} that a higher magnetic field
$\Omega \simeq 1.5 \sqrt{\psi}/2$ than predicted here is needed for
complete stability. That might perhaps be explained by the simulation
potential being more peaked than ${\rm sech}^4(z/4\lambda_D)$ or by
trapped electron distributions differing from the negative-temperature
Maxwellian form used here. But it might also indicate a shortcoming in
the assumption made here that the eigenmode is a pure shift mode.  It
is possible that a distorted or more complicated mode might remain
less stable near the shift mode's marginal stability.  Simulations
with non-Maxwellian background
plasma\citep{Goldman1999,Oppenheim1999,Newman2001a,Oppenheim2001b,Berthomier2002,Lu2008,Wu2010}
have also observed instabilities of different character and shorter
wavelength even above this higher stability limit. However, since
isotropic Maxwellian background simulations do \emph{not} suffer those
``whistler'' or ``streaked'' instabilities, it seems likely that they
are caused by the non-Maxwellian background distribution. Maxwellian
was deliberately chosen here so to be an unequivocally stable
background. The present analysis can readily be extended to address
such cases as anisotropic bi-Maxwellians, but that and other related
extensions are left for future research.

\section*{Acknowledgments}

I am grateful to Chuteng Zhou for our many thoughtful discussions of
electron hole kinematics, and to the anonymous author of a previous
paper who alerted us to the ideas in section 3.1. Xiang Chen helped by
checking some of the algebra. This work was partially supported
by NASA grant NNX16AG82G.

\bibliography{MyAll,hutch18}
\end{document}